\documentclass[usenatbib,useAMS]{mn2e}

\usepackage{amsfonts}
\usepackage{amsmath}
\usepackage{wasysym}
\usepackage{graphicx}

\begin{document}

\label{firstpage}

\title[Top-heavy IMF in UCDs]{A top-heavy stellar initial mass function in starbursts as an explanation for the high mass-to-light ratios of ultra compact dwarf galaxies}

\author[Dabringhausen, Kroupa \& Baumgardt]{
J. Dabringhausen$^{}$\thanks{E-mail: joedab@astro.uni-bonn.de},
P. Kroupa$^{}$\thanks{pavel@astro.uni-bonn.de} and
H. Baumgardt$^{}$\thanks{holger@astro.uni-bonn.de}\\
$^{}$ Argelander-Institut f\"ur Astronomie, Universit\"at Bonn, Auf dem
H\"ugel 71, 53121 Bonn, Germany}

\pagerange{\pageref{firstpage}--\pageref{lastpage}} \pubyear{2008}

\maketitle

\begin{abstract}
It has been shown recently that the dynamical $V$-band mass-to-light ratios of compact stellar systems with masses from $10^6 \, \rmn{M}_{\odot}$ to $10^8 \, \rmn{M}_{\odot}$ are not consistent with the predictions from simple stellar population (SSP) models. Top-heavy stellar initial mass functions (IMFs) in these so-called ultra compact dwarf galaxies (UCDs) offer an attractive explanation for this finding, the stellar remnants and retained stellar envelopes providing the unseen mass. We therefore construct a model which quantifies by how much the IMFs of UCDs would have to deviate in the intermediate-mass and high-mass range from the canonical IMF in order to account for the enhanced $M/L_V$ ratio of the UCDs. The deduced high-mass IMF in the UCDs depends on the age of the UCDs and the number of faint products of stellar evolution retained by them. Assuming that the IMF in the UCDs is a three-part power-law equal to the canonical IMF in the low-mass range and taking 20\% as a plausible choice for the fraction of the remnants of high-mass stars retained by UCDs, the model suggests the exponent of the high-mass IMF to be $\approx 1.6$ if the UCDs are $13 \, \rmn{Gyr}$ old (i.e. almost as old as the Universe) or $\approx 1.0$ if the UCDs are $7 \, \rmn{Gyr}$ old, in contrast to 2.3 for the Salpeter-Massey IMF. If the IMF was as top-heavy as suggested here, the stability of the UCDs might have been threatened by heavy mass loss induced by the radiation and evolution of massive stars. The central densities of UCDs must have been in the range $10^6$-$10^7 \, \rmn{M}_{\odot} \, \rmn{pc}^{-3}$ when they formed with star formation rates of 10-$100 \, \rmn{M}_{\odot} \, \rmn{yr}^{-1}$.
\end{abstract}


\section{Introduction}
\label{Introduction}

Ultra compact dwarf galaxies (UCDs) are stellar systems in which $10^6 \, \rmn{M}_{\odot}$ to $10^8 \, \rmn{M}_{\odot}$ of gas were converted into stars within a volume of some ten pc in diameter \citep{Hil1999, Dri2000, Dri2003, Phi2001, Has2005}. If UCDs are essentially the massive end of the globular cluster sequence \citep{Mie2002,Mie2004,For2008}, then this must have happened within a few Myr, so that the star formation rate would have been 10-100 $\rmn{M}_{\odot} \rmn{yr}^{-1}$. Indeed, the enhancement in $\alpha$-elements, that \citet{Evs2007} found in most of the UCDs they examined, suggests a short time scale for the formation of their stellar populations. Taken together, these properties indicate that UCDs once were among the most extreme star-forming regions in the universe.

A fundamental function underlying star formation is the stellar initial mass function (IMF), $\xi(m)$,
\begin{equation}
dN \propto \xi(m) \, dm,
\end{equation}
where $dN$ is the number of stars with initial masses between $m$ and $m+dm$. The IMF is the parent distribution for the mass functions of stars in star clusters \citep{Kro2003}. These mass functions are subject to statistical scatter \citep{Elm1997,Kro2001} and have an upper mass limit determined by the mass of the gas cloud out of which the star cluster formed \citep{Wei2006}.

One of the most debated questions concerning the IMF is whether it is universal, i.e. independent on the conditions under which star formation takes place. This is not expected from a theoretical point of view. \citet{Ada1996} and \citet{Lar1998} suggest an increase of the characteristic masses of pre-stellar cloud cores with increasing ambient temperature. \citet{Mur1996} discuss interactions of pre-stellar clumps leading to mergers as a process in star formation. Their model predicts an increase of the mean stellar mass with the density of the star-forming region. At the transition from massive globular clusters (GCs) to UCDs (i.e. in the mass range between $10^6$ and $10^7 \, \rmn{M}_{\odot}$), encounters between pre-stellar clumps must have been particularly important. Only about 100 times the diameter of the orbit of Neptune is available for the mean distance between stars in the central parts of some of these high-mass GCs or low-mass UCDs (see fig.~4 in \citealt{Dab2008}, hereafter DHK). If expansion due to mass-loss through gas expulsion and stellar evolution played a role during their youth, then the densities of UCDs would have been even higher at their birth.

On the other hand, all observed \emph{resolved} stellar populations are consistent with having formed with the same IMF. This \emph{canonical} IMF can be formulated as a two-part power law,
\begin{equation}
\xi_{\rmn{c}}(m) = k_i m^{-\alpha_i},
\label{eqIMF}
\end{equation}
with
\begin{eqnarray}
\nonumber \alpha_1 = 1.3,  & \qquad & 0.1  \apprle  \frac{m}{\rmn{M}_{\odot}} < 0.5,\\
\nonumber \alpha_2 = 2.3,  & \qquad & 0.5  \le  \frac{m}{\rmn{M}_{\odot}} \le m_{\rmn{max}},
\end{eqnarray}
where $m_{\rmn{max}}$ is a function of the natal stellar mass of an embedded star cluster at the time when star formation is over and $\xi_{\rmn{c}}=0$ for $m>m_{\rmn{max}}$ \citep{Kro2001,Kro2008}. The factors $k_i$ ensure that the IMF is continuous where the power changes.

During the past years suggestions for the IMF not being universal, but over-abundant in high-mass stars (top-heavy) under extreme conditions, have accumulated for different types of stellar systems. These include galaxies (e.g. \citealt{Bau2005}, \citealt{Nag2005} and \citealt{Dok2008}), the Galactic bulge and centre (e.g. \citealt{Bal2007} and \citealt{Man2007}) and Galactic globular clusters (e.g. \citealt{Dan2004} and \citealt{Pra2006}).

Especially Milky Way globular clusters (MWGCs) have been examined closely. Their stellar mass functions might have been altered strongly by early residual gas expulsion \citep{Mar2008} and stellar and dynamical evolution \citealt{Bau2003,Bor2007,Kru2008a,Kru2008b}), but the observation of individual stars in the MWGCs can still give clues on their IMFs; namely by interpreting the complex patterns of the element abundances in MWGC stars (e.g. the Na-O anti-correlation, see \citealt{Gra2004} for a review on the composition of MWGC stars). These peculiarities are usually taken as evidence for self-enrichment, meaning that the last stars that formed in a particular MWGC contain material that has been processed by stars that formed earlier in the same cluster.

Different theories on how exactly the process of self-enrichment took place have been brought forward: the metal-enrichment in subsequent stellar generations could be caused by the ejecta of massive asymptotic-giant-branch stars, as suggested e.g. by \citet{Dan2004} and \citet{Dan2007}, or by the winds from very massive stars, as suggested e.g. by \citet{Pra2006} and \citet{Dec2007}. Yet both approaches require a top-heavy IMF, although residual gas expulsion from mass-segregated clusters alleviates this need \citep{Dec2008}.

It was shown e.g. in DHK and \citet{For2008} that GCs and UCDs do not constitute two clearly distinguishable populations, if a sample that covers the whole mass interval from GCs to massive UCDs is considered. This suggests a close relation between GCs and UCDs. It therefore seems well possible that the peculiarities in the element abundances that are found for stars in massive MWGCs could as well be present in the even more massive UCDs. But the only nearby objects that may be considered as UCDs and can (like the MWGCs) be resolved into individual stars are $\omega$~Cen and (at least to some extent) G1 in M31. Such observations indeed show the stellar content of these most massive star clusters (or low-mass UCDs) to have a spread of metallicities and ages (e.g. \citealt{Mey2001}, \citealt{Kay2006} and \citealt{Vil2007}).

However, there is an alternative way to set constrains on the IMFs of the UCDs, namely by the comparison with simple stellar population (SSP) models. Various authors thereby found that the UCDs tend to have higher dynamical $V$-band mass-to-light ($M/L_V$) ratios than expected for any possible stellar population that formed with the canonical IMF (\citealt{Has2005,Hil2007,Rej2007}; DHK; \citealt{Mie2008b}).

This result could indicate the presence of non-baryonic dark matter \citep{Has2005,Bau2008b}. However, \citet{Mur2008} argues that both numerical simulations and observations of dwarf spheriodal galaxies hint to dark matter densities that are far too low to influence the dynamics of UCDs. This strengthens the notion that the high $M/L_V$ ratios of the UCDs are the consequence of an IMF different from the canonical one. \citet{Mie2008} discuss an over-abundance of low-mass stars (i.e. stars with high $M/L_V$ ratios) as a possible cause for the high $M/L_V$ ratios of the UCDs. They make testable predictions based on the CO-index \citep{Kro1994}. Complementary to their approach, this contribution is dedicated to top-heavy IMFs as an explanation for the high $M/L_V$ ratio of the UCDs, which is in this scenario the consequence of a large number of remnants from burnt-out stars in them. The possible need for a top-heavy IMF also in the context of the element anti-correlations in massive GCs, as outlined above, makes this approach particularly attractive.

This paper is organised as follows. In Section~\ref{data}, the data sample used in this work is introduced. Section~\ref{stellarpop} describes the model that is constructed for the stellar populations in UCDs. The results suggested by this model for the IMF of intermediate-mass and high-mass stars are presented in Section~\ref{results}. Some implications of these results are discussed in Section~\ref{discussion}. We summarise and conclude in Section~\ref{conclusions}.

\section{The data sample}
\label{data}

The present paper is based on the data of GCs and UCDs compiled in \citet{Mie2008b}, their table~5, because the chosen sample fulfils two requirements necessary for what is done in the present paper:
\begin{enumerate}
 \item{Estimates for the \emph{dynamical mass} have to be available for the objects.}
 \item{Estimates of the global metallicity of the objects have to be possible.}
\end{enumerate}
This sample is currently the largest and most updated sample of its kind. We note however that the results in DHK are qualitatively unchanged, although the present sample has been revised and enhanced compared to the sample they use.

The term 'dynamical mass' refers to a mass estimate that is based on the velocity dispersion of the stars in the stellar system (derived from spectral line widths) and the spatial structure of the stellar system (see \citealt{Hil2007} for details). The mass estimates are therefore independent from the observed total luminosities of the stellar systems.

The metallicities of the stellar systems are of importance for the present paper because of their influence on the luminosity of stellar populations. Knowing them is therefore essential for creating models of stellar populations with a certain $M/L_V$ ratio, which is the focus of the present paper.

Besides newly estimated quantities, table~5 in \citet{Mie2008b} also comprises numbers that are taken from the previous literature, as documented in their paper for the masses but not for the metallicities. Details on the origin of the metallicity estimates for objects with masses $\ge 2 \times 10^6 \, \rmn{M}_{\odot}$ are given in Tab.~\ref{alpha3comp} of our paper. When \citet{Mie2008b} make their own metallicity estimate from the $(V-I)$ colours of the stellar systems they use the relation
\begin{equation}
\rmn{[Fe/H]}=3.27(V-I)-4.50
 \label{eq:FeVI}
\end{equation}
(eq.~4 in \citealt{Kis1998}). This has been done for all objects in their sample with masses $<2 \times 10^6 \, \rmn{M}_{\odot}$, unless the stellar systems are MWGCs for which the metallicities are taken from \citet{Har1996} (private communication with S. Mieske).

Following \citet{Mie2008b}, we take an estimated mass of $\ge 2 \times 10^6 \, \rmn{M}_{\odot}$ as an easy-to-handle criterion to categorise a compact stellar system as a UCD instead of a GC. This mass marks quite well the transition from objects with GC-like properties to objects with UCD-like properties (\citealt{Mar2005}; \citealt{Mie2008b}; DHK), including the on average distinctively higher $M/L_V$ ratios of the more massive objects. Note that the two-body relaxation time exceeds a Hubble time for systems larger than $2 \times 10^6 \, \rmn{M}_{\odot}$ \citep{Mie2008}, which has been proposed as the defining property to distinguish galaxies from star clusters (\citealt{Kro1998}; DHK).

\section{A Model for the stellar populations of the UCDs}
\label{stellarpop}

We now construct a model for the stellar populations of the UCDs under the assumption that the deviations of their $M/L_V$ ratios from the theoretical expectation for the $M/L_V$ ratio of a stellar population with the canonical IMF are caused by an IMF that varies for intermediate-mass and high-mass stars. The actual shape of the IMF in the UCDs cannot be specified from resolved stellar populations so far. The purpose of the following can therefore only be to give an idea by how much the IMF must deviate from the canonical IMF in order to account for the mismatch between observations and theoretical expectations for the $M/L_V$ ratio of the UCDs.

\subsection{The model ingredients}
\label{model}

The problem of modelling a stellar population with a $M/L_{V}$ ratio equal to an observed value can be formulated as
\begin{equation}
\frac{M_{\rmn{m}}}{L_{\rmn{m}}}- \Upsilon_V =0,
\label{eqML1}
\end{equation}
where $M_{\rmn{m}}$ is the total mass of the model population, $L_{\rmn{m}}$ is its luminosity in the $V$-band and $\Upsilon_{V}$ is the observed $M/L_{V}$ ratio of a stellar system. $M_{\rmn{m}}$ and $L_{\rmn{m}}$ depend on various parameters, such as the assumed age of the population, the shape of its IMF and the chosen model for stellar and cluster evolution. $L_{\rmn{m}}$ additionally depends on the metallicity. These dependencies will be formulated below, along with the assumptions that are made for the model presented here.

\subsubsection{The IMF}
\label{IMF}

The IMFs of the UCDs are connected to their present-day mass functions in the simplest way possible, because of their median two-body relaxation times, $t_{\rmn{rh}}$, which are of the order of a Hubble time or larger (DHK; \citealt{Mie2008b}). The timescale on which a stellar system dissolves depends on the tidal field strength, but can be expected to be many $t_{\rmn{rh}}$, so that the stellar populations of UCDs are practically unaltered by dynamical evolution. This stands in contrast to GCs, whose $t_{\rmn{rh}}$ are much shorter and therefore can have experienced significant dynamical evolution since their formation (also see Section~\ref{single}).

We introduce a family of IMFs for the model stellar populations of the UCDs:
\begin{equation}
\xi_{\rmn{pl}}(m) = k_i m^{-\alpha_i},
\label{IMFf}
\end{equation}
with
\begin{eqnarray}
\nonumber \alpha_1 = 1.3,  & \qquad & 0.1  \le  \frac{m}{\rmn{M}_{\odot}} < 0.5,\\
\nonumber \alpha_2 = 2.3,  & \qquad & 0.5  \le  \frac{m}{\rmn{M}_{\odot}} < 1, \\
\nonumber \alpha_3 \, \in \, \mathbb{R}, & \qquad & 1 \le  \frac{m}{\rmn{M}_{\odot}} \le m_{\rmn{max}},
\end{eqnarray}
where $m_{\rmn{max}}$ is the upper mass limit for stars. These IMFs will be referred to as the 'three-part power-law IMFs'. They are equal to the canonical IMF except for their slope above $ 1 \, \rmn{M}_{\odot}$. We assume that the UCDs have formed with a three-part power-law IMF.

Upper mass limits of $100 \, \rmn{M}_{\odot}$ and $150 \, \rmn{M}_{\odot}$ are considered. The upper mass limit of $100 \, \rmn{M}_{\odot}$ equals the upper mass limit assumed in the simple stellar population (SSP) models which are used in this paper (see Section~\ref{SSP}). These are the same stellar population models DHK took as a reference when they found that the $M/L_{V}$ ratios of a significant majority of the UCDs tends to be higher than model predictions for the canonical IMF. This mass limit is however not in agreement with the upper mass limit for stars in very massive star clusters given by \citet{Wei2004}, \citet{Oey2005} and \citet{Fig2005}, which is close to $150 \, \rmn{M}_{\odot}$. Therefore this more realistic upper mass limit is considered as well. It turns out that the results are affected surprisingly little by the upper mass limit of the IMF (Figs.~\ref{alphaUps} and~\ref{alphaM} below).

Note that the lower mass limit of the IMF neglects the existence of brown dwarfs. This is probably unproblematic, since \citet{Thi2007} showed that a combined mass function of brown dwarfs and stars shows a discontinuity. In the case that the low-mass IMFs of the UCDs are comparable to the ones in Galactic open star clusters (as assumed to be the case here), their results suggest that brown dwarfs contribute only a few percent to the total mass of the UCDs.

The formulation of the IMF in UCDs given in eq.~\ref{IMFf} attributes a possibly enhanced $M/L_V$ ratio of a UCD solely to a top-heavy IMF, i.e. to a large population of stellar remnants that would have to be expected in such a case. Note however that the assumption of a bottom-heavy, Salpeter-Massey-like IMF in UCDs is currently an equally valid approach to explain their $M/L_V$ ratios (cf. figure~10 in DHK). Observations to test the hypothesis of a bottom-heavy IMF in UCDs using a method proposed in \citet{Mie2008} are underway.

\subsubsection{Simple stellar population models}
\label{SSP}

A simple stellar population (SSP) is defined as a population of stars of the same age and metallicity. Various authors have set up grids of models of such populations, e.g. \citet{Bru2003} and \citet{Mar2005}. The alteration of the stellar mass function due to dynamical evolution is not considered in these grids; only stellar evolution changes the mass spectrum of the stars in the model populations.

The most closely examined object in the sample of UCDs used here, $\omega$~Cen, is known to have several stellar sub-populations of different ages and metallicities, i.e. $\omega$~Cen is not a SSP (e.g. \citealt{Hil2000,Hil2004,Vil2007}). Still, the sub-populations in $\omega$ Cen can all be characterised as old and metal-poor. Taking $\omega$~Cen in this sense as representative for the UCDs, we assume that their stellar populations are composed of different sub-populations, but that these sub-populations are \emph{similar} enough to describe each UCD as a single SSP for the purpose of this paper. Also note that stellar-encounter-driven dynamical evolution is negligible in the UCDs (DHK). A disagreement between the SSP models and the observations can in this light be interpreted as being caused by assuming the wrong IMF.

The SSP models of \citet{Bru2003} and \citet{Mar2005} differ by the stellar evolutionary models used to calculate the luminosity of the modelled population as well as the total mass assumed for this population. \citet{Bru2003} assume a somewhat higher mass-loss rate for the stellar populations (\citealt{Mar2005}, in particular her figure~22), while the luminosities they get from the stellar models they use are lower. In effect, the estimates for the $M/L$ ratios by \citet{Bru2003} are similar to the ones by Maraston (2005, her figure~24). However, considering the predictions for the $M/L_V$ ratios of old populations, the estimates by \citet{Bru2003} are about 20\% lower than the ones by \citet{Mar2005}. Note that this cannot be accounted for by the different formulations for the canonical IMF these authors use since they turn out to be nearly identical (figure~8 in DHK). In fact, \citet{Bru2003} find that the stellar mass of a 10~Gyr old population is 52\% of the initial stellar mass for the canonical IMF they use, while it would have been 54\% if they had used the same formulation of the IMF as \citet{Mar2005} does. The reminder of the difference in the $M/L_V$ ratio of an old stellar population must thus be the consequence of the different stellar evolutionary models used and different assumptions regarding the remnant masses (also see DHK and \citealt{Mie2008b}). As a compromise between the two sets of SSP models, we follow the approach by \citet{Mie2008b} and take the mean of the predictions from \citet{Bru2003} and \citet{Mar2005} as the reference for a comparison to the observations in UCDs.

We consider ages of $7 \, \rmn{Gyr}$ and $13 \, \rmn{Gyr}$ for the UCDs, since these values are at the limits of the ages expected for them. An age of $7 \, \rmn{Gyr}$ would be consistent with the intermediate age for the Fornax UCDs suggested in \citet{Mie2006} and \citet{Mie2008b}. Note that assuming even younger ages would increase the discrepancy between the observed $M/L_{V}$ ratio and the model predictions. Ages higher than $13 \, \rmn{Gyr}$ are excluded by the estimates for the age of the universe ($13.73^{+0.16}_{-0.15} \, \rmn{Gyr}$; \citealt{Spe2007}).

The turn-off mass from the main sequence for a population of coeval stars, $m_\rmn{to}$, marks quite well the stellar mass above which stars of that population have already evolved into stellar remnants. It is $\approx1 \, \rmn{M}_{\odot}$ for a $\approx10 \, \rmn{Gyr}$ stellar population. Since stellar evolution is slow for old stars, $m_{\rmn{to}}=1 \, \rmn{M}_{\odot}$ is a reasonably good approximation for a $7 \, \rmn{Gyr}$ old SSP as well as for a $13 \, \rmn{Gyr}$ old SSP.

The contribution of the stellar remnants to the $V$-band luminosity of the UCDs is small and therefore neglected in this paper. The luminosity, $L_{\rmn{m}}$, of a modelled stellar population is thus insensitive to the degree of top-heavyness of the IMF, since the IMF is only allowed to vary in a mass range where the stars have evolved after $\approx 10 \, \rmn{Gyr}$. The masses of the stars that have not evolved yet are assumed to be distributed in concordance with the canonical IMF. Thus, $L_{\rmn{m}}$ can be determined using the SSP models from \citet{Bru2003} and \citet{Mar2005} with the canonical IMF.

We note that by this approach the influence of binary systems on stellar evolution is neglected.

\subsubsection{The initial-to-final-mass relation for stars}

In order to find an explicit formulation of $M_{\rmn{m}}$ in eq.~(\ref{eqML1}), a formulation of the masses of evolved stars as a function of their initial masses is needed. This function, called the initial-to-final-mass relation, $m_\rmn{rem}(m)$, allows to calculate the total mass of an evolved SSP from its IMF for a given age. Using the three-part power-law IMFs from Section~\ref{IMF}, the integral that has to be solved in this calculation reads
\begin{equation}
 M_{\rmn{m}}=\int^{m_\rmn{max}}_{0.1} m_{\rmn{rem}}(m) \xi_{\rmn{pl}}(m) \, dm,
\label{eqMrem2}
\end{equation}
where $m$ is the stellar initial mass in $\rmn{M}_{\odot}$. The limits of the integration are set by the lower and the upper initial mass limit for stars.

The initial-to-final-mass relation used in this paper is specified in the following.

For stars with initial masses $m<m_{\rmn{to}}$, $m_{\rmn{rem}}=m$ is assumed, i.e. the mass loss of main-sequence stars is neglected.

Stars with initial masses of $m_{\rmn{to}}<m<8 \, \rmn{M}_{\odot}$ are assumed to have evolved into white dwarfs (WDs), in concordance with the mass limit given by \citet{Koe1996}. \citet{Kal2008} find, performing a weighted least-squares fit of a linear function to data based on observations of WDs in star clusters,
\begin{equation}
m_{\rmn{rem}}=(0.109 \pm 0.007) \frac{m}{\rmn{M}_{\odot}}+(0.394 \pm 0.025),
\label{eqWD}
\end{equation}
for a relation between the mass of WDs and the initial mass of their progenitors, where $m$ is the stellar initial mass in $\rmn{M}_{\odot}$. This relation is adopted in this paper.

Stars initially more massive than $8 \, \rmn{M}_{\odot}$ but less massive than $\approx25 \, \rmn{M}_{\odot}$ are predicted to evolve into neutron stars (NSs) with a remarkably narrow mass spread (cf. figures 12 and 16 in \citealt{Woo2002}). This is observationally supported by \citet{Tho1999}, who find the mass-distribution of pulsars (i.e. observable NSs) in their data sample to be consistent with a Gaussian distribution with a mean of $1.35 \, \rmn{M}_{\odot}$ and a width of $0.04 \, \rmn{M}_{\odot}$. Thus, in this paper $1.35 \rmn{M}_{\odot}$ is adopted for the masses of all stellar remnants with initial masses between $8 \, \rmn{M}_{\odot}$ and $25 \, \rmn{M}_{\odot}$.

Stars with initial masses above $25 \, \rmn{M}_{\odot}$ are generally thought to be the progenitors of stellar-mass black holes (BHs). However, the theoretical predictions for the masses of their remnants are not only strongly dependent on metallicity, but also on the assumptions on how the evolution of such stars proceeds (see figures~12 and~16 in \citealt{Woo2002}). Figure~12 in \citet{Woo2002} might suggest that the case of the higher remnant masses is the more appropriate choice for low-metallicity environments such as GCs and UCDs. However, the masses of observationally confirmed BHs lie all in a range that is covered by assuming that the remnants of very high-mass stars only have 10\% of the initial mass of their progenitors \citep{Cas2006}. In our paper, we thus assume that stars with $m>25 \, \rmn{M}_{\odot}$ evolve into BHs that have either 10\% or 50\% of the mass of their progenitor stars, but the emphasis is on the case with the less massive BHs because of the observational support for their existence.

Note that BHs formed through single-star evolution differ from NSs in mass, but not in the processes that precede their creation. NSs and BHs are both compact remnants that emerge from the core collapse and SN explosion of a massive star.

To summarise, the complete initial-to-final-mass function, $m_{\rmn{rem}}$ used here is
\begin{equation}
m_{\rmn{rem}}=
\begin{cases}
\frac{m}{\rmn{M}_{\odot}}, & \frac{m}{\rmn{M}_{\odot}} < \frac{m_{\rmn{to}}}{\rmn{M}_{\odot}}, \\
0.109 \, \frac{m}{\rmn{M}_{\odot}}+0.394, & \frac{m_{\rmn{to}}}{\rmn{M}_{\odot}} \le \frac{m}{\rmn{M}_{\odot}} < 8, \\
1.35, &  8  \leq  \frac{m}{\rmn{M}_{\odot}} < 25, \\
0.1\frac{m}{\rmn{M}_{\odot}} \ \rmn{or} \ 0.5\frac{m}{\rmn{M}_{\odot}}, &  25  \leq \frac{m}{\rmn{M}_{\odot}} \le m_{\rmn{max}},
\end{cases}
\label{eqMrem1}
\end{equation}
where $m_{\rmn{to}}$ denotes the turn-off mass and $m_{\rmn{max}}$ the upper initial mass limit for stars (eqs.~\ref{eqIMF} and~\ref{IMFf}). Inserting eq.~(\ref{eqMrem1}) into eq.~(\ref{eqMrem2}) and carrying out the integration on the right hand side of eq.~(\ref{eqMrem2}) yields the mass of all stars and stellar remnants as a function of only the high-mass IMF-slope, $\alpha_3$, if $m_{\rmn{to}}$ (i.e. age) and $m_{\rmn{max}}$ are specified. The terms resulting from this integration for initial stellar masses above $m_{\rmn{to}}$ are written down explicitly in Appendix~\ref{Equations}.

\subsubsection{Normalised mass-to-light ratios}
\label{normML}

The metallicities estimated for the UCDs usually do not coincide with the grid points of the SSP models by \citet{Bru2003} and \citet{Mar2005}. It is therefore necessary to find interpolation formulae that describe the metallicity dependency of the $M/L_{V}$ ratio in the models (which actually is a dependency of the luminosity on metallicity). This can be done by fitting exponential functions of the form
\begin{equation}
 F|_{Z}=F([Z/\rmn{H}])=\left(a^{[Z/\rmn{H}]+b}+c \right) \, \frac{\rmn{M}_{\odot}}{\rmn{L}_{\odot}},
\label{MetML}
\end{equation}
to the data from \citet{Bru2003} and \citet{Mar2005}, where $[Z/\rmn{H}]$ is the metallicity. The best-fitting parameters $a$, $b$ and $c$ found in a least-squares fit to the models used in this paper are listed in Table \ref{tabFit}. The excellent agreement of this type of function to the models from \citet{Bru2003} and \citet{Mar2005} is demonstrated in fig.~9 in DHK.

\begin{table}
\caption{Fit parameters for the metallicity-dependent interpolation formula for $\Upsilon_V$ to the data from SSP models with the canonical IMF. BC indicates SSP models from \citet{Bru2003} and M SSP models from \citet{Mar2005}.}
\label{tabFit}
\centering
\vspace{2mm}
\begin{tabular}{llll}
\hline
&&&\\[-10pt]
Model & $a$ & $b$ & $c$ \\ [2pt]
\hline
&&&\\[-10pt]
BC, 7 Gyr & $3.29$ & $0.12$ & $1.05$\\
M, 7 Gyr & $3.26$ & $0.22$ & $1.24$\\
BC, 13 Gyr & $3.48$ & $0.55$ & $1.71$\\
M, 13 Gyr & $3.46$ & $0.79$ & $1.88$\\
\hline
\end{tabular}
\end{table}

The reference relation that is taken to describe the metallicity dependency of the $M/L_V$ ratio for a SSP with a certain age and with the canonical IMF is the mean of the corresponding relations derived from the SSP models from \citet{Bru2003} and \citet{Mar2005} (cf. \citealt{Mie2008b}). The ratio between the observed $M/L_{V}$ ratio for a UCD and the result from the reference relation at the appropriate metallicity is a measure for the discrepancy between the observed value and the theoretical prediction. It is convenient for the purpose here to multiply these values by the prediction of the reference relation for the $M/L_{V}$ ratio at Solar metallicity. These quantities will be referred to as normalised $M/L_{V}$ ratios, $\Upsilon_{V\rmn{,n}}$,
\begin{equation}
 \Upsilon_{V\rmn{,n}}=\frac{\Upsilon_{V}}{F_{BC}|_{Z}+F_{M}|_{Z}}\times (F_{BC}|_{\rmn{Z}_{\odot}}+F_{M}|_{\rmn{Z}_{\odot}}),
\end{equation}
where a subscript BC indicates that the parameters $a$, $b$ and $c$ correspond to a SSP model from \citet{Bru2003} and the subscript M indicates that the parameters $a$, $b$ and $c$ correspond to a SSP model from \citet{Mar2005} (for the same age).

\begin{figure*}
\includegraphics[scale=0.72]{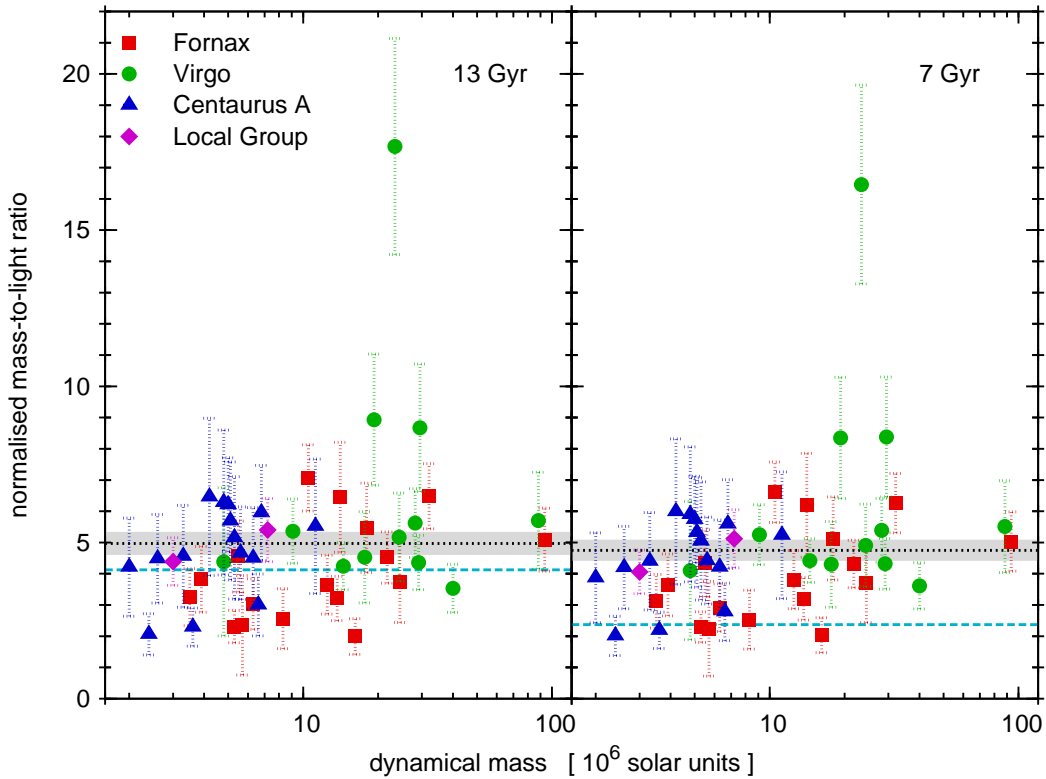}
\caption{Normalised $M/L_{V}$ ratios, $\Upsilon_{V\rmn{,n}}$, of the stellar systems collected in table~5 in \citet{Mie2008}, provided their dynamical mass is estimated to be $2\times 10^6 \, \rmn{M}_{\odot}$ or more. Thus, the figure only shows objects that are UCDs according to the definition used in this paper. The ages assumed for them are either $13 \, \rmn{Gyr}$ (left panel) or $7 \, \rmn{Gyr}$ (right panel). The dashed horizontal lines indicate $\Upsilon_{V\rmn{,n}}$ for a SSP that formed with the canonical IMF and is of the age that is assumed for the UCDs in the according panels. The dotted horizontal lines correspond to the mean of the $\Upsilon_{V\rmn{,n}}$ of all UCDs in the sample and the shaded areas indicate the uncertainty given to this value. These numbers are used to estimate the high mass-slope of the UCDs (see Section~\ref{independent}).}
\label{MLratios}
\end{figure*}

Using these values for $\Upsilon_{V\rmn{,n}}$, eq.~(\ref{eqML1}) can be rewritten as
\begin{equation}
\frac{M_{\rmn{m}}}{L_{\rmn{m}}|_{\rmn{Z}_{\odot}}}- \Upsilon_{V\rmn{,n}}=0.
\label{eqML2}
\end{equation}
$L_{\rmn{m}}|_{\rmn{Z}_{\odot}}$ is thereby no longer a metallicity-dependent variable, but is fixed to the value the reference relation predicts for Solar metallicity and  thereby only depends on the age assumed in the model and the amplitude of the factors $k_{\rmn{i}}$ in the IMF. The metallicity dependency is shifted into the transformation from the observed $M/L_{V}$ ratio of the UCD to $\Upsilon_{V\rmn{,n}}$. The $\Upsilon_{V\rmn{,n}}$ values are noted in Tab.~\ref{alpha3comp} and shown in Fig.~\ref{MLratios}. Their uncertainties have been propagated from the errors of the observed dynamical $M/L_{V}$ ratios and the errors of the metallicity estimates.

The numerical value of $L_m$ is calculated from the secondary condition that the prediction for $\Upsilon_{V\rmn{,n}}$ from the SSP models should correspond to a stellar population with the canonical IMF with $m_{\rmn{max}}=100 \, \rmn{M}_{\odot}$ and a full population of remnants (i.e. a stellar population as in the SSP models). For this, $m_{\rmn{rem}}$ as given in eq.~(\ref{eqMrem1}) is used, adopting the case that the black hole masses, $m_{\rmn{BH}}$, are 10\% the stellar initial masses ($m_{\rmn{BH}}=0.1 m$).

There is evidence that GCs usually have $\alpha$-enrichments, $[\alpha / \rmn{Fe}]$, of $0.3 \, \rmn{dex}$ \citep{Car1996}. \citet{Evs2007} find that the same $[\alpha / \rmn{Fe}]$ is also typical for the UCDs in the Virgo cluster they examine. On the other hand, \citet{Mie2007} find that a number of UCD candidates is consistent with having Solar $[\alpha / \rmn{Fe}]$, which is why \citet{Mie2008b} adopt Solar $[\alpha / \rmn{Fe}]$ for all stellar systems in their study. However, assuming a super Solar $[\alpha / \rmn{Fe}]$ is the more careful choice in the context of the present paper, since it attributes more of a possibly enhanced $M/L_V$ ratio in UCDs to metallicity effects. As in DHK, we therefore adopt $[\alpha / \rmn{Fe}]=0.3 \, \rmn{dex}$ for all GCs and UCDs and estimate their metallicities, $[Z/\rmn{H}]$, from their iron abundances, [Fe/H]. This is done using the relation
\begin{equation}
[Z/\rmn{H}]=\rmn{[Fe/H]}+0.94 \, [\alpha/\rmn{Fe}]
\label{eq:alpha}
\end{equation}
taken from \citet{Tho2003}. Consequently, the $[Z/\rmn{H}]$ used for calculating the $\Upsilon_{V\rmn{,n}}$ are $0.28 \, \rmn{dex}$ higher than the [Fe/H] and the $\Upsilon_{V\rmn{,n}}$ in this paper are thereby slightly lower than the ones in \citet{Mie2008b}.

The assumed age turns out to be almost irrelevant for the $\Upsilon_{V\rmn{,n}}$ calculated for the individual stellar systems. However, the assumed age does have a strong impact on the $\Upsilon_{V\rmn{,n}}$ predicted by the SSP-models (see also DHK).

\subsubsection{The fate of the processed material and the stellar remnants in the UCDs}
\label{processed}

In order to have an influence on the dynamics of a stellar system, the stellar remnants that form in it have to remain bound to it. This can be assumed to be the case for the WDs in the UCDs, since WDs inherit the peculiar velocities of their progenitor stars and two-body encounter driven mass loss is negligible for the UCDs (see Section~\ref{SSP}).

Unlike the case with WDs, stellar evolution has a direct impact on the velocity distribution of NSs. It is well established that many pulsars move with high peculiar velocities, which they must have obtained somehow in their formation out of their progenitor stars \citep{Woo1987,Lyn1994}. \citet{Lyn1994} give the mean pulsar birth velocity as $450 \pm 90 \, \rmn{km} \, \rmn{s}^{-1}$. Since the processes that lead to the formation of BHs through single-star evolution are the same as the ones that precede the formation of NSs, the BHs should also receive kicks.

The UCDs have velocity dispersions of $\apprle 50 \rmn{km} \, \rmn{s}^{-1}$, which suggests escape velocities of the order of $\apprle 100 \rmn{km} \, \rmn{s}^{-1}$. Thus, the peculiar velocities of most NSs and BHs should be high enough to leave the UCDs. On the other hand, NSs are known to populate GCs, which suggests that also the UCDs are able to retain some fraction of these objects.

Most of the matter processed in intermediate-mass and high-mass stars is reinserted as gas and dust into the interstellar medium during stellar evolution. Its fate is therefore crucial for the developement and consequently the $M/L_V$ ratio of a stellar system.

There are in general three possibilities for what can happen to this material. If it remains inside the cluster, it can (at least in principle) simply accumulate (and thereby emit almost no radiation in the $V$-band) or it can be used up in the formation of subsequent stellar populations. Alternatively, the gas can be driven out of the UCDs, e.g. by type~I~SNe or by the ram pressure caused by the movement of the UCD through the intergalactic medium.

Gas and dust originating from intermediate-mass stars has a good chance to stay inside the UCDs, since these stars form in their final stage planetary nebulae that expand with moderate velocities ($\approx20 \, \rmn{km \, s^{-1}}$; see e.g. \citealt{Ges2003}). These velocities are too low for the matter to leave a star cluster with a deep potential well immediately. This makes massive AGB stars attractive progenitors for a second generation of stars, as proposed in \citet{Dan2004} and \citet{Dan2007}.

However, massive stars evolve into SN and thereby release $\approx10^{50} \, \rmn{erg}$ per Solar unit of initial mass of the progenitor star (cf. fig.~1 in \citealt{Nom2006}). This clearly exceeds the binding energy of a star to a UCD. Material originating from these stars will therefore easily escape from the stellar system, unless the kinetic energy of the gas from the SN explosion is dissipated (e.g. by the interaction with primordial gas or the collision of expanding gas envelopes from different SNe with one another). The gas density and the holding time inside the UCDs might become long enough for the gas to cool and to collapse, as discussed in \citet{Ten2007}.

Note however that neither self-enrichment by massive AGB stars nor self-enrichment with SN ejecta can explain the multiple stellar populations in $\omega$~Cen, since both scenarios act on a time scale of $\apprle200 \, \rmn{Myr}$, whereas the age difference between the different stellar populations in $\omega$~Cen is a few Gyr \citep{Hil2000, Hil2004,Vil2007}.

The essence of this is that the current knowledge on the evolution of the UCDs does not allow solid conclusions on the composition of the UCDs. We therefore consider six different compositions of the UCDs for which we estimate the high-mass IMF-slope:

\begin{enumerate}
 \item Out of all material from burnt out stars, only WDs are retained by the UCDs. This can be taken as the lower limit for the amount of matter that stays inside the UCDs since the UCDs are nearly unaffected  by dynamical evolution (cf. Section~\ref{SSP}).
 \item 20\% of the compact remnants from stars initially more massive than $8 \, \rmn{M}_{\odot}$ are retained by the UCDs. The remnant masses of stars with $m>25 \, \rmn{M}_{\odot}$, $m_{\rmn{BH}}$, are assumed to be 10\% of the initial mass of their progenitors, $m_{\rmn{BH}}=0.1 m$. The NS and BH retention rate of 20\% is an arbitrarily chosen value, but this scenario might still be close to a realistic one. On the one hand it allows for some NSs in the UCDs as observed for GCs, but on the other hand it also takes into account that the observed velocity dispersion of pulsars is high by estimating the fraction of retained NSs to be low.
 \item As scenario (ii), but with $m_{\rmn{BH}}=0.5 m$.
 \item All stellar remnants are retained by the UCDs and $m_{\rmn{BH}}=0.1 m$. Such a population, where all stars and stellar remnants, but not the ejecta from stars are considered, is assumed in the SSP models.
 \item As scenario (iv), but with $m_{\rmn{BH}}=0.5 m$.
 \item The UCDs were gas-free after star-formation ceased in them, but all material that was processed in burnt-out stars is retained by the UCDs and star formation with the gaseous component of this matter is somehow inhibited. Of all the models considered here, this is the one where stars contribute the least to the total mass of the UCD (consisting of stars, remnants and possibly gas). Note however that the hydrodynamic calculations by \citet{Ten2007} suggest that such a scenario is unlikely because the gas accumulating in the UCDs due to stellar evolution will more likely either leave the UCDs or collapse into new stars.
\end{enumerate}

Interstellar gas and all remnants in the UCDs are considered not to contribute to the light of the UCDs. In other words, the very high $M/L_{V}$ ratios of these components of the UCDs are taken to be infinity.

\section{results}
\label{results}

\begin{figure*}
\includegraphics[scale=0.72]{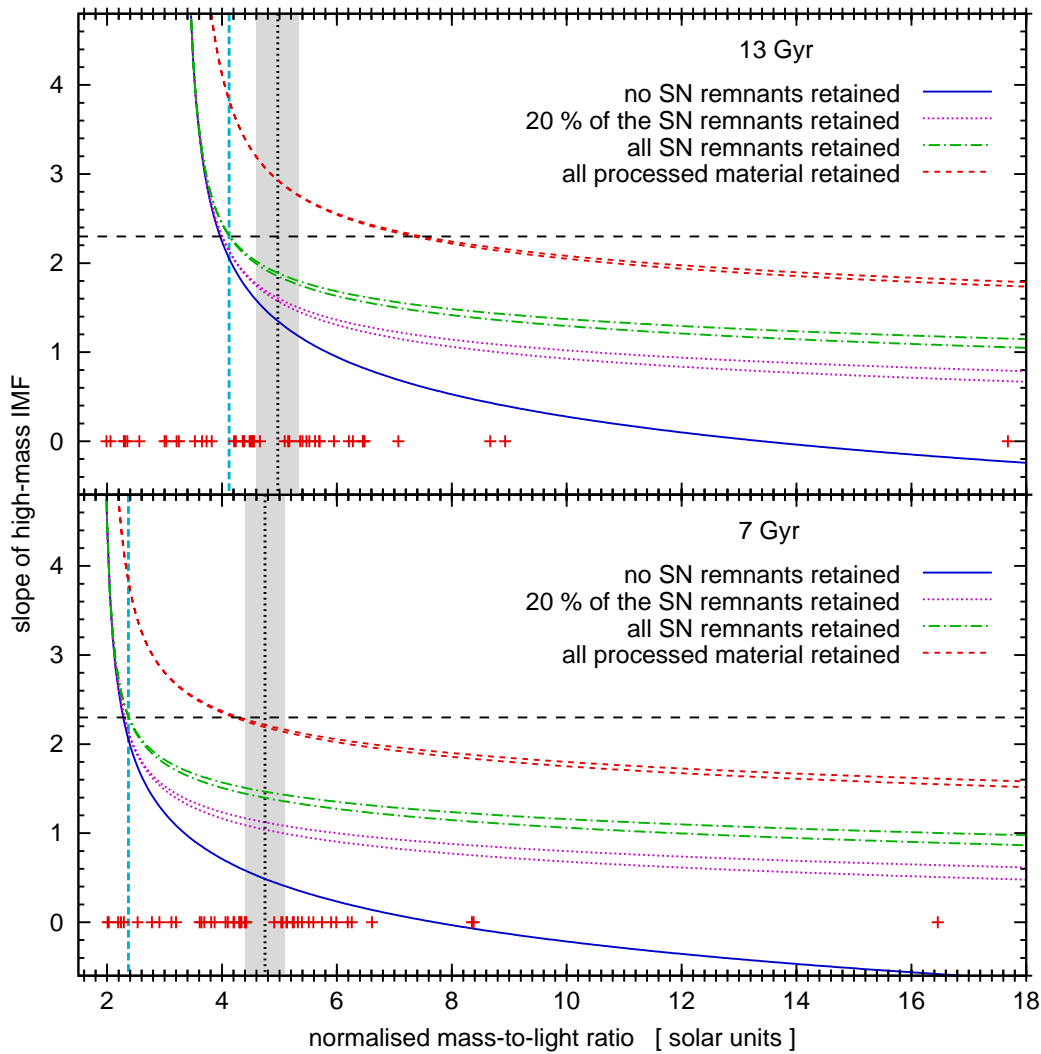}
\caption{The high-mass slope, $\alpha_3$, against the normalised mass-to-light ratio, $\Upsilon_{V,n}$, as implied by solving eq.~(\ref{eqML2}) for a three-part power-law IMF (see eq.~\ref{IMFf} and the equations in the appendix). The curves are for a $13 \, \rmn{Gyr}$ old SSP (upper panel) and for a $7 \, \rmn{Gyr}$ old SSP (lower panel). The different styles of the curves correspond to different assumptions on how much processed matter (with extremely high $M/L_{V}$ ratio) besides WDs is retained by the UCDs (from the bottom to the top curves in each panel): no remnants of massive stars; 20\% of the remnants of massive stars; all remnants of massive stars; all material processed by burnt-out stars. Two curves of the same style indicate different assumptions for the upper mass limit of the IMF for the same assumption on the matter retained in the UCDs: $100 \, \rmn{M}_{\odot}$ (lower curve) and $150 \, \rmn{M}_{\odot}$ (upper curve). The dashed horizontal lines indicate in each panel the canonical high-mass IMF index, $\alpha_3=2.3$. Its intersections with the curves show the $\Upsilon_{V\rmn{,n}}$ which the canonical IMF would imply for a particular remnant population.  The $\Upsilon_{V\rmn{,n}}$ of the individual UCDs are shown as crosses at the bottom of each panel. The dashed vertical lines indicate $\Upsilon_{V\rmn{,n}}$ for a SSP that formed with the canonical IMF and is of the age that is assumed for the UCDs in the according panels. The dotted vertical lines correspond to the mean of the $\Upsilon_{V\rmn{,n}}$ of all UCDs in the sample and the shaded areas indicate the uncertainty given to this value (see Section~\ref{independent}). The intersections of a vertical line with the curves show that $\alpha_3$ corresponding to a particular $\Upsilon_{V\rmn{,n}}$ for the different assumptions on the retained remnant population. In this Figure, the remnants of stars with $m>25 \, \rmn{M}_{\odot}$ are assumed to have masses of 10\% of the initial mass of their progenitors whereever this is relevant.}
\label{alphaUps}
\end{figure*}

The value for the high-mass IMF slope, $\alpha_3$, implied by a given normalised $M/L_V$ ratio, $\Upsilon_{V\rmn{,n}}$, for the assumptions on the stellar populations in the UCDs specified in Section~\ref{stellarpop} can be calculated from eq.~(\ref{eqML2}) using the Newton-Raphson root-finding method.

There is a lower limit for the $\Upsilon_{V\rmn{,n}}$ that leads to a solution for eq.~(\ref{eqML2}), because the lowest $\Upsilon_{V\rmn{,n}}$ that can be realised within the model is the one for a stellar population whose IMF is cut off at $1 \, \rmn{M}_{\odot}$ (i.e. $\alpha_3=\infty$). If the age of the UCDs is assumed to be $13 \, \rmn{Gyr}$, the individual $\Upsilon_{V\rmn{,n}}$ of a number of UCDs is actually below that limit.

Close to that lower limit, $\alpha_3$ increases rapidly with decreasing $\Upsilon_{V\rmn{,n}}$, implying a steep high-mass IMF (Fig.~\ref{alphaUps}). In this range, the solutions to eq.~(\ref{eqML2}) become degenerate for the different assumptions on how much mass is retained by the UCDs. That is because a steep high-mass IMF means few high-mass stars and it is therefore not decisive for the $\Upsilon_{V\rmn{,n}}$ of an old stellar system how much matter from those stars is retained.

The $\alpha_3$ of the canonical IMF is in the regime where the relation between $\Upsilon_{V\rmn{,n}}$ and $\alpha_3$ is already close to being degenerate for different assumptions on the NS and BH retention rate. Only assuming all matter processed in burnt-out stars remains inside the UCDs without forming new stars would lead to a distinctively higher $\Upsilon_{V\rmn{,n}}$ for $\alpha_3=2.3$. In other words, the predictions of the models from \citet{Bru2003} and \citet{Mar2005} for the $\Upsilon_{V\rmn{,n}}$ for a SSP with the canonical IMF (shown as the horizontal dashed line in Fig.~\ref{MLratios} and as the vertical dashed line in Fig.~\ref{alphaUps}) depend, except for extreme cases, only weakly on the fate assumed for the material processed in massive stars.

Comparing solutions of eq.~(\ref{eqML2}) for the same $\Upsilon_{V\rmn{,n}}$ and remnant retention rate, but  for upper mass limits of $100 \, \rmn{M}_{\odot}$ and $150 \, \rmn{M}_{\odot}$, reveals that the remnants of very massive stars do not play a decisive role for the $\alpha_3$ that are obtained, as illustrated in Figs.~\ref{alphaUps} and \ref{alphaM}. This finding may be surprising, since the total mass of the remnants of high-mass stars is a function of the exponent $\alpha_3$  (see Appendix~\ref{Equations}). This mass must therefore increase dramatically with increasing $\alpha_3$ above some critical value for $\alpha_3$.

The results of solving eq.~(\ref{eqML2}) if a  remnant retention rate of 20\% and $m_{\rmn{BH}}=0.1 m$ is assumed are noted in Table~\ref{alpha3comp}. However, for many individual UCDs solutions do not exist if a high age is assumed for them (i.e. their $\Upsilon_{V\rmn{,n}}$ is clearly below the prediction from the SSP models for a canonical IMF), and the uncertainties are large in any case. On the other hand, application of the Pearson test for the goodness of fit  (cf. \citealt{Bha1977} and DHK) on the 46 UCDs from \citet{Mie2008b} shows that the actual distribution of the $\Upsilon_{V\rmn{,n}}$ of the UCDs in the sample is highly unlikely if their individual $\Upsilon_{V\rmn{,n}}$ scatters equally to both sides of the prediction for the $\Upsilon_{V\rmn{,n}}$ of a SSP with the canonical IMF (less than 1\% if the age of the UCDs is assumed to be at its maximum, $13 \, \rmn{Gyr}$, and much less than 0.5\% if the age of the UCDs is assumed to be $7 \, \rmn{Gyr}$). The properties of the \emph{sample} of UCDs therefore imply an IMF that deviates from the canonical IMF (provided that they do not contain non-baryonic DM), such as a three-part power-law IMF with $\alpha_3<2.3$.

The emphasis in this paper is therefore on constraining likely values for the high-mass IMF slopes of the UCDs from the properties of the whole sample of UCDs and different subsamples thereof. It is decisive for this to know whether the $\Upsilon_{V\rmn{,n}}$ of the UCDs are correlated with their mass, $M$, and to quantify this correlation if there is one (Section \ref{alpha3Mass}). If such a dependency is found, the dependency of $\alpha_3$ on $\Upsilon_{V\rmn{,n}}$ can be translated into a dependency of $\alpha_3$ on $M$.

\begin{table*}
\caption{Normalised $M/L_{V}$ ratios, $\Upsilon_{V\rmn{,n}}$, of the UCDs for assumed ages of $7 \, \rmn{Gyr}$ and $13 \, \rmn{Gyr}$, and the high-mass slopes, $\alpha_3$, these $\Upsilon_{V\rmn{,n}}$ suggest if 20 \% of the remnants of massive stars are retained by the UCDs, BHs have 10\% of the initial mass of their progenitor stars and the upper mass limit of the IMF is $m_{\rmn{max}}=100 \, \rmn{M}_{\odot}$. The contents of the columns are the following: Column~1: The object identification (as in \citealt{Mie2008b}, table~5), Column~2: The projected half-light radius of the UCD, Column~3: The estimate for the iron-abundance, Column~4: The mass of the UCD, Column~5: Its $\Upsilon_{V\rmn{,n}}$ based on the models by \citet{Bru2003} and \citet{Mar2005} for a $7 \, \rmn{Gyr}$ old SSP with the canonical IMF, Column~6: The estimate for $\alpha_3$ based on the value for $\Upsilon_{V\rmn{,n}}$ in Column~5, Columns~7 and~8: As Columns~5 and~6 respectively, but for an assumed age of the UCD of $13 \, \rmn{Gyr}$. The superscript numbers in Column 3 indicate the origin of the [Fe/H] estimate: 1: \citet{Mie2008b}, 2: \citet{Has2005}, 3: \citet{Mey2001}, 4: \citet{Har1996}. A superscript * indicates that [Fe/H] was not obtained from colour indices, but from line indices or the properties of the resolved stellar population of the stellar object (private communication with S. Mieske). $\Upsilon_{V\rmn{,n}}$ is estimated using $[Z/\rmn{H}]$, which is $\approx 0.3 \, \rmn{dex}$ higher than the corresponding [Fe/H] due to the assumed $\alpha$-enhancement of the stellar systems (see Section~\ref{normML}). Dots in Columns~5 and~8 indicate where no solution for eq.~(\ref{eqML2}) is found under the given assumptions. The canonical IMF would have $\alpha_2=\alpha_3=2.3$ (Salpeter-Massey index, eq.~\ref{eqIMF}).}
\vspace{2mm}
\centering
\begin{tabular}{lrrr@{\hspace{0.2mm}}lrrrr}
\hline
&&&&&&&\\[-10pt]
Name  &  $M$ & $r_{\rmn{e}}$ & \multicolumn{2}{r}{[Fe/H]} & $\Upsilon_{V\rmn{,n}}$ &  $\alpha_{3}$ & $\Upsilon_{V\rmn{,n}}$ & $\alpha_{3}$ \\[2pt]
      &	     &	             &     &   			  &  (7 Gyr)	           & (7 Gyr)	   & (13 Gyr)               & (13 Gyr)\\
      &  [$10^6 \, \rmn{M}_{\odot}$]  & [pc] & & & [$\rmn{M}_{\odot} \, \rmn{L}_{\odot, V}^{-1}$] & & [$\rmn{M}_{\odot} \, \rmn{L}_{\odot , V}^{-1}$] &\\

\hline

&&&&&&&\\[-10pt]
F-7       & 10.5 & 14.9 & $-1.$&$3^{1}$ & $6.61\pm 0.97$ & $0.86_{-0.06}^{+0.08}$ & $7.07\pm 1.06$ & $1.15_{-0.10}^{+0.15}$\\
UCD1      & 32.1 & 22.4 & $-0.$&$7^{1}$ & $6.26\pm 0.95$ & $0.88_{-0.07}^{+0.09}$ & $6.48\pm 1.04$ & $1.23_{-0.12}^{+0.20}$\\
F-9       & 14.1 &  9.1 & $-0.$&$8^{1}$ & $6.19\pm 1.66$ & $0.89_{-0.11}^{+0.18}$ & $6.45\pm 1.76$ & $1.23_{-0.19}^{+0.46}$\\
UCD5      & 18.0 & 31.2 & $-1.$&$2^{1}$ & $5.13\pm 1.33$ & $0.99_{-0.12}^{+0.22}$ & $5.47\pm 1.43$ & $1.42_{-0.25}^{+0.82}$\\
F-19      & 93.6 & 89.7 & $-0.$&$4^{1*}$ & $5.03\pm 0.95$ & $1.00_{-0.10}^{+0.15}$ & $5.09\pm 1.01$ & $1.53_{-0.24}^{+0.65}$\\
F-34      &  5.5 &  4.9 & $-0.$&$9^{1}$ & $4.34\pm 1.07$ & $1.10_{-0.14}^{+0.27}$ & $4.55\pm 1.14$ & $1.77_{-0.40}^{+\dots}$\\
UCD2      & 21.8 & 32.1 & $-0.$&$9^{1*}$ & $4.31\pm 0.76$ & $1.11_{-0.11}^{+0.17}$ & $4.52\pm 0.82$ & $1.79_{-0.34}^{+1.22}$\\
F-6       & 12.5 &  7.3 & $0.$&$2^{1*}$ & $3.81\pm 0.93$ & $1.21_{-0.17}^{+0.35}$ & $3.82\pm 1.05$ & $2.65_{-1.03}^{+\dots}$\\
F-24      & 24.5 & 29.5 & $-0.$&$4^{1*}$ & $3.69\pm 1.25$ & $1.24_{-0.22}^{+0.76}$ & $3.73\pm 1.29$ & $2.91_{-1.35}^{+\dots}$\\
F-53      &  3.9 &  4.4 & $-0.$&$9^{1}$ & $3.64\pm 0.99$ & $1.25_{-0.19}^{+0.49}$ & $3.65\pm 0.94$ & $3.22_{-1.47}^{+\dots}$\\
F-5       & 13.7 &  5.0 & $-0.$&$3^{1}$ & $3.20\pm 0.68$ & $1.39_{-0.20}^{+0.49}$ & $3.25\pm 0.91$ & $\dots$\\
F-51      &  3.5 &  4.2 & $-0.$&$8^{1}$ & $3.12\pm 0.86$ & $1.43_{-0.26}^{+0.98}$ & $3.21\pm 0.71$ & $\dots$\\
F-17      &  6.3 &  3.3 & $-0.$&$8^{1}$ & $2.91\pm 0.76$ & $1.54_{-0.30}^{+1.33}$ & $3.03\pm 0.81$ & $\dots$\\
F-12      &  8.3 & 10.3 & $-0.$&$4^{1*}$ & $2.53\pm 0.94$ & $1.87_{-0.57}^{+\dots}$ & $2.56\pm 0.96$ & $\dots$\\
F-22      &  5.3 & 10.0 & $-0.$&$4^{1*}$ & $2.29\pm 0.49$ & $2.32_{-0.69}^{+\dots}$ & $2.35\pm 1.59$ & $\dots$\\
F-11      &  5.7 &  3.6 & $-0.$&$9^{1}$ & $2.24\pm 1.52$ & $2.47_{-1.25}^{+\dots}$ & $2.31\pm 0.52$ & $\dots$\\
F-1       & 16.2 & 23.1 & $0.$&$0^{1*}$ & $2.03\pm 0.56$ & $3.95_{-2.15}^{+\dots}$ & $1.99\pm 0.57$ & $\dots$\\
S999      & 23.4 & 19.1 & $-1.$&$4^{2}$ &$16.46\pm 3.18$ & $0.51_{-0.06}^{+0.07}$ &$17.68\pm 3.46$ & $0.67_{-0.07}^{+0.09}$\\
S417      & 29.5 & 14.4 & $-0.$&$7^{2}$ & $8.38\pm 1.92$ & $0.75_{-0.08}^{+0.12}$ & $8.93\pm 2.10$ & $0.99_{-0.11}^{+0.19}$\\
S928      & 19.3 & 21.8 & $-1.$&$3^{2}$ & $8.35\pm 1.94$ & $0.75_{-0.08}^{+0.12}$ & $8.67\pm 2.04$ & $1.01_{-0.12}^{+0.20}$\\
VUCD7     & 88.3 & 96.8 & $-0.$&$7^{1*}$ & $5.51\pm 1.47$ & $0.95_{-0.12}^{+0.21}$ & $5.70\pm 1.55$ & $1.37_{-0.23}^{+0.73}$\\
VUCD1     & 28.2 & 11.3 & $-0.$&$8^{1*}$ & $5.39\pm 1.02$ & $0.96_{-0.09}^{+0.13}$ & $5.62\pm 1.10$ & $1.38_{-0.19}^{+0.41}$\\
S314      &  9.1 &  3.2 & $-0.$&$5^{2}$ & $5.25\pm 0.96$ & $0.98_{-0.09}^{+0.13}$ & $5.36\pm 1.03$ & $1.45_{-0.21}^{+0.48}$\\
VUCD4     & 24.3 & 22.0 & $-1.$&$0^{1*}$ & $4.91\pm 1.32$ & $1.02_{-0.13}^{+0.24}$ & $5.17\pm 1.41$ & $1.50_{-0.29}^{+1.31}$\\
S490      & 14.5 &  3.6 & $0.$&$2^{2}$ & $4.42\pm 0.70$ & $1.09_{-0.10}^{+0.14}$ & $4.53\pm 1.46$ & $1.78_{-0.48}^{+\dots}$\\
VUCD5     & 29.1 & 17.9 & $-0.$&$4^{1*}$ & $4.32\pm 0.81$ & $1.10_{-0.11}^{+0.18}$ & $4.36\pm 0.87$ & $1.90_{-0.42}^{+\dots}$\\
VUCD6     & 17.7 & 14.8 & $-1.$&$0^{1*}$ & $4.30\pm 1.37$ & $1.11_{-0.17}^{+0.42}$ & $4.24\pm 0.75$ & $2.01_{-0.44}^{+\dots}$\\
H8005     &  4.8 & 28.1 & $-1.$&$3^{2}$ & $4.10\pm 2.21$ & $1.14_{-0.27}^{+\dots}$ & $4.38\pm 2.37$ & $1.89_{-0.70}^{+\dots}$\\
VUCD3     & 40.0 & 18.7 & $0.$&$0^{2}$ & $3.61\pm 0.74$ & $1.26_{-0.16}^{+0.31}$ & $3.53\pm 0.77$ & $3.98_{-2.02}^{+\dots}$\\
HCH99-2  &  4.2 & 11.4 & $-1.$&$5^{1}$ & $5.99\pm 2.33$ & $0.91_{-0.15}^{+0.34}$ & $6.46\pm 2.52$ & $1.23_{-0.24}^{+1.16}$\\
HGHH92-C21&  4.8 &  7.0 & $-1.$&$2^{1}$ & $5.90\pm 2.16$ & $0.91_{-0.15}^{+0.31}$ & $6.28\pm 2.32$ & $1.26_{-0.24}^{+1.10}$\\
VHH81-C5  &  5.0 & 10.0 & $-1.$&$6^{1}$ & $5.74\pm 1.38$ & $0.93_{-0.11}^{+0.17}$ & $6.21\pm 1.50$ & $1.27_{-0.18}^{+0.42}$\\
HGHH92-C1 &  6.8 & 24.0 & $-1.$&$2^{1}$ & $5.59\pm 1.42$ & $0.94_{-0.11}^{+0.19}$ & $5.95\pm 1.52$ & $1.31_{-0.21}^{+0.54}$\\
HGHH92-C17&  5.1 &  5.7 & $-1.$&$3^{1}$ & $5.32\pm 1.75$ & $0.97_{-0.15}^{+0.30}$ & $5.69\pm 1.89$ & $1.37_{-0.27}^{+1.33}$\\
HCH99-18  & 11.2 & 13.7 & $-1.$&$0^{1}$ & $5.23\pm 2.03$ & $0.98_{-0.17}^{+0.41}$ & $5.52\pm 2.15$ & $1.41_{-0.32}^{+\dots}$\\
HGHH92-C11&  5.3 &  7.8 & $-0.$&$5^{1}$ & $5.05\pm 1.89$ & $1.00_{-0.17}^{+0.41}$ & $5.15\pm 1.96$ & $1.51_{-0.36}^{+\dots}$\\
HCH99-15  &  5.6 &  5.9 & $-1.$&$0^{1}$ & $4.42\pm 1.39$ & $1.09_{-0.17}^{+0.38}$ & $4.66\pm 1.48$ & $1.71_{-0.43}^{+\dots}$\\
HGHH92-C29&  3.3 &  6.9 & $-0.$&$7^{1}$ & $4.40\pm 1.56$ & $1.09_{-0.18}^{+0.49}$ & $4.56\pm 1.63$ & $1.77_{-0.49}^{+\dots}$\\
HGHH92-C7 &  6.3 &  7.5 & $-1.$&$3^{1}$ & $4.21\pm 1.51$ & $1.12_{-0.19}^{+0.57}$ & $4.50\pm 1.62$ & $1.80_{-0.52}^{+\dots}$\\
HGHH92-C22&  2.6 &  3.8 & $-1.$&$2^{1}$ & $4.20\pm 1.32$ & $1.13_{-0.18}^{+0.43}$ & $4.48\pm 1.42$ & $1.82_{-0.49}^{+\dots}$\\
HCH99-16  &  2.0 & 12.1 & $-1.$&$9^{1}$ & $3.87\pm 1.44$ & $1.19_{-0.22}^{+0.82}$ & $4.21\pm 1.57$ & $2.04_{-0.69}^{+\dots}$\\
HGHH92-C23&  6.6 &  3.3 & $-1.$&$5^{1}$ & $2.78\pm 0.92$ & $1.63_{-0.39}^{+\dots}$ & $3.00\pm 0.99$ & $\dots$\\
HGHH92-C6 &  3.6 &  4.4 & $-0.$&$9^{1}$ & $2.19\pm 0.58$ & $2.67_{-1.03}^{+\dots}$ & $2.29\pm 0.61$ & $\dots$\\
VHH81-C3  &  2.4 &  4.4 & $-0.$&$6^{1}$ & $2.01\pm 0.63$ & $4.28_{-2.53}^{+\dots}$ & $2.06\pm 0.66$ & $\dots$\\
G1        &  7.2 &  3.0 & $-1.$&$0^{3}$ & $5.12\pm 0.93$ & $0.99_{-0.09}^{+0.14}$ & $5.40\pm 1.01$ & $1.44_{-0.20}^{+0.44}$\\
$\omega$~Cen & 3.0  &  8.0 & $-1.$&$6^{4}$ & $4.06\pm 0.70$ & $1.15_{-0.11}^{+0.18}$ & $4.39\pm 0.76$  & $1.88_{-0.37}^{+1.44}$ \\
\hline
\end{tabular}
\label{alpha3comp}
\end{table*}

\subsection{Does $\alpha_3$ depend on mass?}
\label{alpha3Mass}

\begin{figure*}
\includegraphics[scale=0.72]{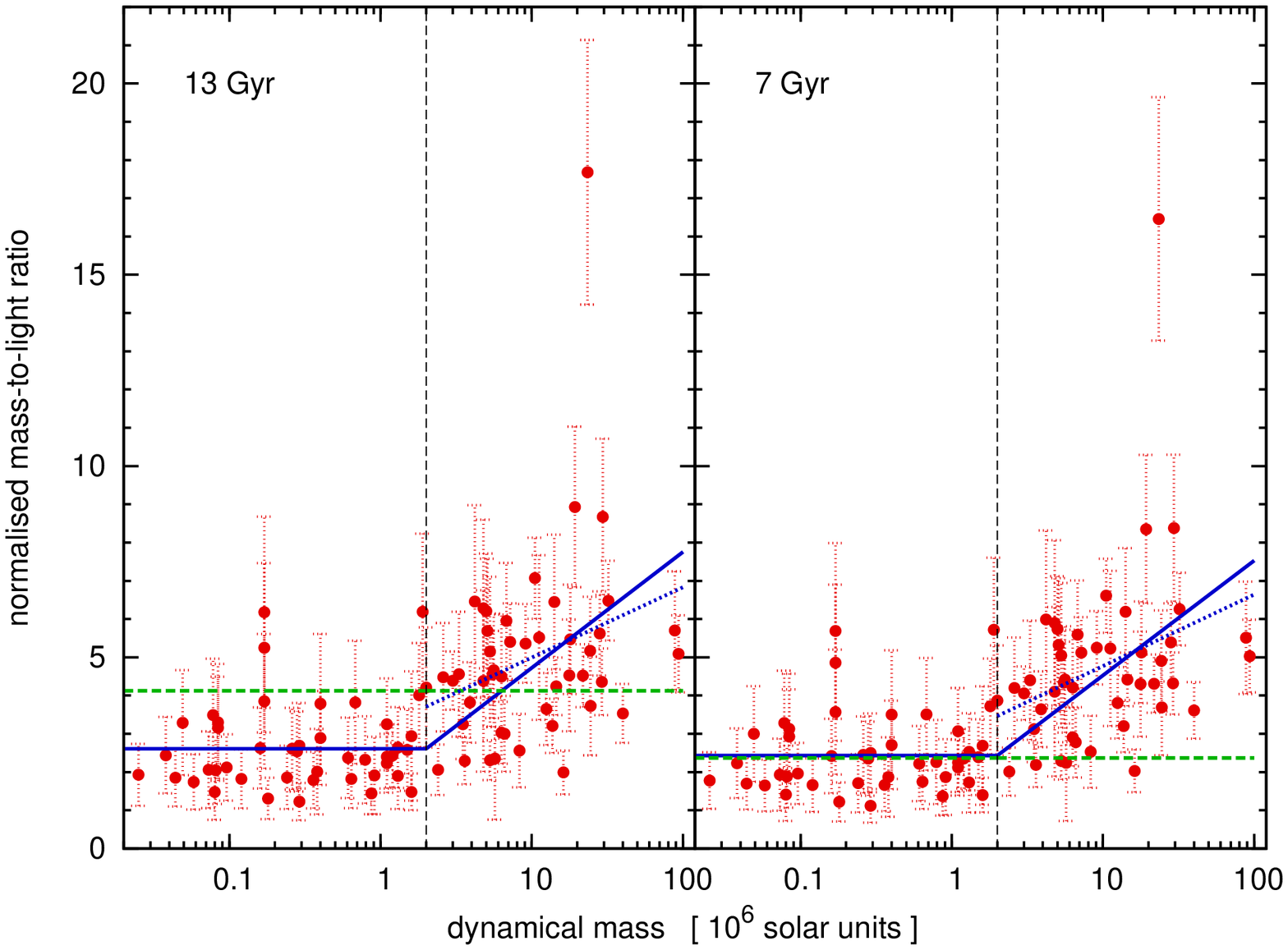}
\caption{Normalised $M/L_{V}$ ratios, $\Upsilon_{V\rmn{,n}}$, of all objects listed in table~5 in \citet{Mie2008b}. The assumed ages for them are either $13 \, \rmn{Gyr}$ (left panel) or $7 \, \rmn{Gyr}$ (right panel). Contrary to Fig. \ref{MLratios}, this figure also shows stellar systems with dynamical masses less than $2\times 10^6 \, \rmn{M}_{\odot}$, i.e. stellar systems that are not UCDs but GCs according to the definition used in this paper. Objects considered as GCs are seperated from objects considered as UCDs by the thin, dashed vertical line in each panel. The solid line indicates $\Upsilon_{V\rmn{,n}}(M)$, a function that describes the systematic increase of the average $\Upsilon_{V\rmn{,n}}$ with mass for the case that GCs and UCDs are a single population. If GCs and UCDs are separate populations, $\Upsilon_{V\rmn{,n}}(M)$ of the UCDs is represented by the dotted line that starts at $2 \times 10^6 \, \rmn{M}_{\odot}$. Note that in this case the uncertainty to the slope is very high and is therefore not significant. The dashed horizontal line indicates the prediction for $\Upsilon_{V\rmn{,n}}$ of an SSP with the canonical IMF.}
\label{MUps}
\end{figure*}

At present, it is unclear whether UCDs are the most massive GCs (e.g. \citealt{Mie2002,Mie2004,For2008}) or whether UCDs and GCs are different populations (e.g. \citealt{Dri2004,Goe2008}). However, the answer to this question has implications on how a dependency of $\Upsilon_{V\rmn{,n}}$ on $M$, $\Upsilon_{V\rmn{,n}}(M)$, has to be formulated for GCs and UCDs. An appropriate formulation of $\Upsilon_{V\rmn{,n}}(M)$ as a representation for the typical $\Upsilon_{V\rmn{,n}}$ of objects with a given mass would be a single, continuous function in the first case, but different functions for GCs and UCDs in the second case.

The MWGCs, which make up most of the GCs in the sample used here (tab.~5 in \citealt{Mie2008b}), show no evidence for a bulk-dependency of $\Upsilon_{V}$ with $M$ \citep{McL2000}. Therefore, the mean $\Upsilon_{V\rmn{,n}}$ of the GCs in the data sample, $\overline{\Upsilon_{\rmn{GC}}}$, is adopted for $\Upsilon_{V\rmn{,n}}(M)$ in this mass range. Thus, for $M< 2\times 10^6 \, \rmn{M}_{\odot}$, $\Upsilon_{V\rmn{,n}}(M)=2.43 \pm 0.16 \, \rmn{M}_{\odot} \, \rmn{L}_{\odot ,V}^{-1}$ if the assumed age is $7 \, \rmn{Gyr}$ and $\Upsilon_{V\rmn{,n}}(M)=2.61 \pm 0.18 \, \rmn{M}_{\odot} \, \rmn{L}_{\odot ,V}^{-1}$ if the assumed age is $13 \, \rmn{Gyr}$ (uncertainties are one-sigma values).

The uncertainties of the data for the UCDs leaves many options for an appropriate formulation of $\Upsilon_{V\rmn{,n}}(M)$ for them. We choose
\begin{equation}
\Upsilon_{V\rmn{,n}}(M)=\left( A\left[\log_{10} \left(\frac{M}{\rmn{M}_{\odot}}\right)-\log_{10}(2\times 10^6)\right]+B\right)\frac{\rmn{M}_{\odot}}{\rmn{L}_{\odot}}
\label{eqMUps}
\end{equation}
for $M > 2\times 10^6 \, \rmn{M}_{\odot}$, where $M$ is in Solar units and $A$ and $B$ are parameters which are either fixed by a secondary condition or determined by a least-squares fit. Note that weighting the uncertainties when fitting is not advisable in this case, as it would cause an unwanted bias. This becomes evident by considering two stellar systems with the same mass and uncertainty of the mass, but different luminosities. The uncertainty of a luminosity measurement is negligible compared to the uncertainty of a mass estimate. The uncertainty of the $M/L$ ratio is thus higher for the stellar system with the higher $M/L$ ratio, even if the parameter that induces this uncertainty is the same for both systems. The parameters $A$ and $B$ are therefore determined with equal weight to every measurement and the uncertainties of $A$ and $B$ are estimated only from the scatter of the data.

In order to constrain $\Upsilon_{V\rmn{n}}(M)$ for the UCDs in the case that UCDs and GCs are two distinct populations, $A$ and $B$ in eq.~(\ref{eqMUps}) are left as free parameters for the fit. The best-fitting parameters are $A=1.84 \pm 0.89$ and $B=3.71 \pm 0.70$ if the UCDs are assumed to be $7 \, \rmn{Gyr}$ old and $A=1.87 \pm 0.81$ and $B=3.47 \pm 0.64$ if the UCDs are assumed to be $13 \, \rmn{Gyr}$ old. This may hint at a systematic increase of the $\Upsilon_{V\rmn{,n}}$ of the UCDs with $M$, but the significance of this result ($\approx 2 \sigma$) is not high enough to allow definite conclusions\footnote{We mention that weighting the UCDs by the uncertainties leads qualitatively to the same results, although the best-fitting values for $A$ and $B$ und their uncertainties are slightly lower.}. This finding is consistent with \citet{Mie2008b}, who performed a similar test but only for the UCDs in Fornax.

If UCDs are the most massive GCs, $\Upsilon_{V\rmn{,n}}(M)$ is expected to be continuous at $M=2\times 10^6 \, \rmn{M}_{\odot}$, which in this paper is taken to be the mass that separates GCs and UCDs (see Section \ref{data}). For this case, $\Upsilon_{V\rmn{,n}}(M)$ of the UCDs is therefore estimated by setting $B$ in eq.~(\ref{eqMUps}) to the numerical value of $\overline{\Upsilon_{\rmn{GC}}}$ in Solar units and leaving only $A$ as a free parameter to be determined in the fit. The result is $A=3.00 \pm 0.42$ if the UCDs are assumed to be $7 \, \rmn{Gyr}$ old and $A=3.04 \pm 0.46$ if the UCDs are assumed to be $13 \, \rmn{Gyr}$ old. In this case, the increase of the $\Upsilon_{V\rmn{,n}}$ of the UCDs with their mass is highly significant.

$\Upsilon_{V\rmn{,n}}(M)$ is plotted in Fig.~\ref{MUps} together with the data for the GCs and the UCDs.

\subsection{Constraining $\alpha_3$ from the whole sample of UCDs}
\label{sample}

\subsubsection{UCDs and GCs as independent populations}
\label{independent}

\begin{table*}
\caption{Estimates for the most likely values of $\alpha_3$ for different assumptions concerning age and remnant population of the UCDs, as detailed in Section~\ref{independent}. The first column specifies the supposed remnant population of the UCDs ($m_{\rmn{max}}=100 \, \rmn{M}_{\odot}$). For the stars more massive than $25 \, \rmn{M}_{\odot}$, the cases of them forming BHs with 10\% their initial mass, $m_{\rmn{BH}}=0.1 m$, or 50\% their initial mass, $m_{\rmn{BH}}=0.5 m$, are considered whereever this makes a difference. The second column displays the $\alpha_3$ corresponding to the mean of the $\Upsilon_{V\rmn{,n}}$ of the UCDs and the uncertainties to $\alpha_3$ calculated from the uncertainties to the mean $\Upsilon_{V\rmn{,n}}$. This can be taken as a convenient number to quantify the high-mass IMF slope of the UCDs as a class of objects. The numbers in Columns~3 to~5 have the same meaning as the numbers in Column~2, but for different subsamples of UCDs. The subsamples are chosen by the larger structures the UCDs are bound to, namely the Fornax Cluster (Column~3), the Virgo Cluster with S999 (Column~4) and without S999 (Column~5) and Centaurus~A (Column~6).}
\centering
\vspace{2mm}
\begin{tabular}{llllll}
\hline
&&&&&\\[-10pt]
& All & Fornax & Virgo & Virgo & Centaurus~A \\
&     &        & (with S999) & (without S999) & \\
\hline
&&&&&\\[-10pt]
Model & $\overline{\alpha_3}$ & $\overline{\alpha_3}$ & $\overline{\alpha_3}$ & $\overline{\alpha_3}$ & $\overline{\alpha_3}$\\
\hline \hline
&&&&&\\[-10pt]
\multicolumn{6}{c}{assumed age of 13 Gyr}\\
\hline \hline
&&&&&\\[-10pt]
no SN remnants & $1.35_{-0.17}^{+0.23}$ & $2.10_{-0.42}^{+0.90}$ & $0.81_{-0.23}^{+0.37}$ & $1.11_{-0.18}^{+0.25}$ & $1.49_{-0.21}^{+0.29}$ \\
20\% of the SN remnants, $m_{\rmn{BH}}=0.1 m$ & $1.57_{-0.12}^{+0.17}$ & $2.17_{-0.35}^{+0.84}$ & $1.22_{-0.13}^{+0.23}$ & $1.41_{-0.12}^{+0.17}$ & $1.68_{-0.15}^{+0.23}$ \\
20\% of the SN remnants, $m_{\rmn{BH}}=0.5 m$ & $1.78_{-0.09}^{+0.13}$ & $2.26_{-0.29}^{+0.76}$ & $1.50_{-0.11}^{+0.19}$ & $1.65_{-0.09}^{+0.14}$ & $1.86_{-0.12}^{+0.18}$ \\
all SN remnants, $m_{\rmn{BH}}=0.1 m$ & $1.85_{-0.10}^{+0.14}$ & $2.33_{-0.29}^{+0.73}$ & $1.56_{-0.12}^{+0.20}$ & $1.72_{-0.10}^{+0.14}$ & $1.94_{-0.12}^{+0.18}$ \\
all SN remnants, $m_{\rmn{BH}}=0.5 m$ & $2.11_{-0.08}^{+0.11}$ & $2.50_{-0.23}^{+0.61}$ & $1.86_{-0.10}^{+0.18}$ & $2.00_{-0.08}^{+0.12}$ & $2.18_{-0.10}^{+0.15}$ \\
all processed material & $2.93_{-0.18}^{+0.26}$ & $3.90_{-0.59}^{+1.55}$ & $2.43_{-0.18}^{+0.32}$ & $2.69_{-0.16}^{+0.25}$ & $3.08_{-0.22}^{+0.36}$ \\
\hline \hline
&&&&&\\[-10pt]
\multicolumn{6}{c}{assumed age of 7 Gyr}\\
\hline \hline
&&&&&\\[-10pt]
no SN remnants & $0.49_{-0.08}^{+0.09}$ & $0.73_{-0.12}^{+0.15}$ & $0.19_{-0.15}^{+0.20}$ & $0.36_{-0.10}^{+0.11}$ & $0.56_{-0.09}^{+0.11}$ \\
20\% of the SN remnants, $m_{\rmn{BH}}=0.1 m$ & $1.04_{-0.04}^{+0.05}$ & $1.17_{-0.07}^{+0.09}$ & $0.88_{-0.08}^{+0.10}$ & $0.97_{-0.05}^{+0.07}$ & $1.09_{-0.05}^{+0.06}$ \\
20\% of the SN remnants, $m_{\rmn{BH}}=0.5 m$ & $1.34_{-0.04}^{+0.04}$ & $1.46_{-0.06}^{+0.07}$ & $1.21_{-0.07}^{+0.09}$ & $1.28_{-0.04}^{+0.05}$ & $1.38_{-0.04}^{+0.04}$ \\
all SN remnants, $m_{\rmn{BH}}=0.1 m$ & $1.40_{-0.04}^{+0.05}$ & $1.52_{-0.06}^{+0.08}$ & $1.25_{-0.07}^{+0.10}$ & $1.33_{-0.05}^{+0.06}$ & $1.43_{-0.05}^{+0.05}$ \\
all SN remnants, $m_{\rmn{BH}}=0.5 m$ & $1.72_{-0.04}^{+0.04}$ & $1.82_{-0.05}^{+0.07}$ & $1.59_{-0.06}^{+0.09}$ & $1.66_{-0.04}^{+0.05}$ & $1.75_{-0.04}^{+0.05}$ \\
all processed material & $2.19_{-0.05}^{+0.07}$ & $2.36_{-0.09}^{+0.12}$ & $2.00_{-0.09}^{+0.13}$ & $2.10_{-0.06}^{+0.08}$ & $2.25_{-0.07}^{+0.08}$ \\
\hline
\end{tabular}
\label{alpha3}
\end{table*}

\begin{figure*}
\includegraphics[scale=0.72]{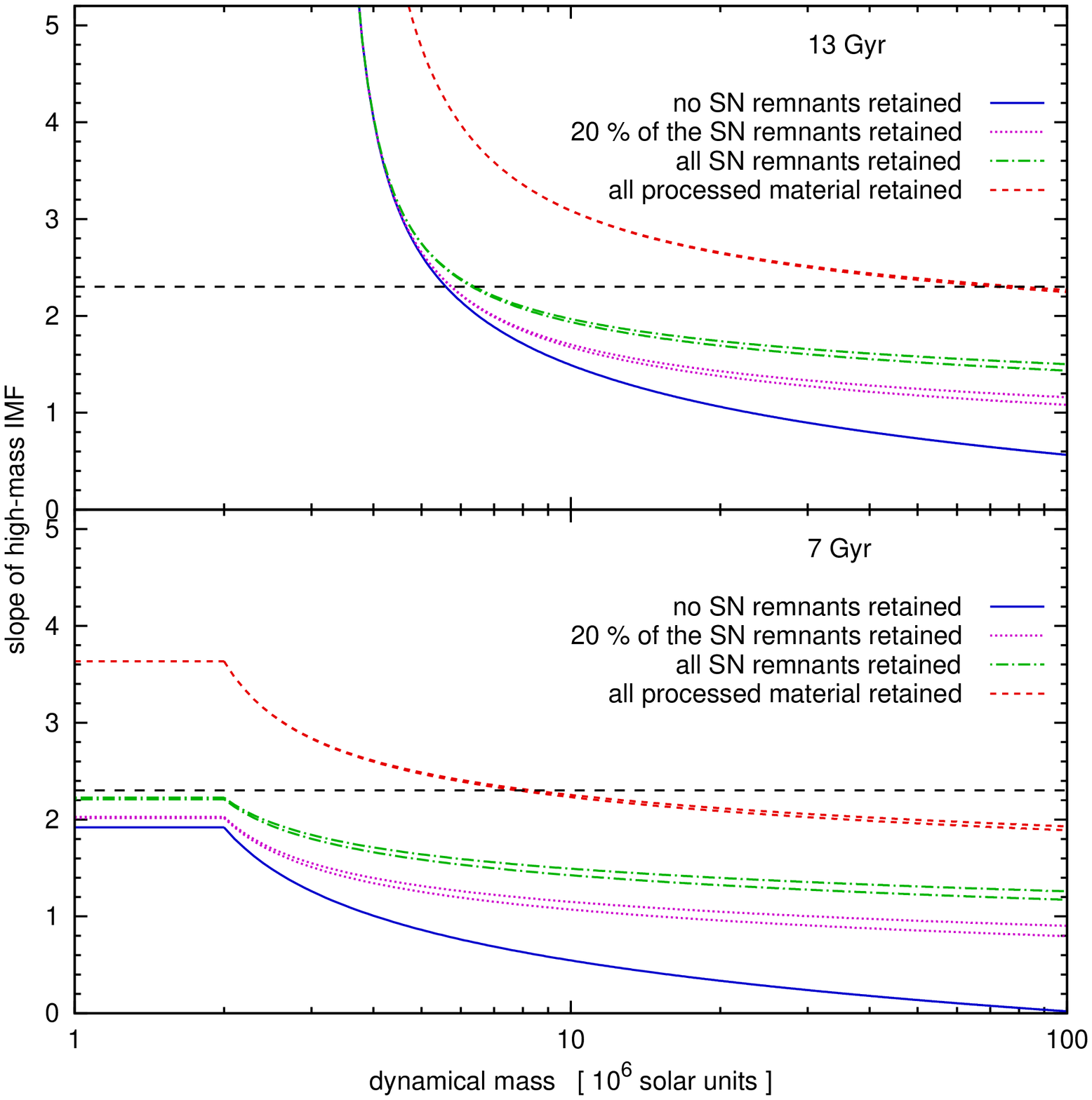}
\caption{The high-mass slope, $\alpha_3$, in dependency of mass $M$, as implied by solving eq.~(\ref{eqML2}) for the values of $\Upsilon_{V\rmn{,n}}(M)$ at the according $M$ and for a three-part power-law IMF (see eq.~\ref{IMFf} and the equations in the Appendix). The curves are for a $13 \, \rmn{Gyr}$ old SSP (upper panel) and for a $7 \, \rmn{Gyr}$ old SSP (lower panel). The different styles of the curves correspond to different assumptions on how much processed matter (with very high $M/L_{V}$ ratio) besides WDs is retained by the UCDs (from the bottom to the top curves in each panel): no remnants of massive stars; 20\% of the remnants of massive stars; all remnants of massive stars; all material processed by burnt-out stars. Two lines of the same style indicate different assumptions for the upper mass limit of the IMF for the same assumption on the matter retained in the UCDs: $100 \, \rmn{M}_{\odot}$ (lower line) and $150 \, \rmn{M}_{\odot}$ (upper line). The dashed horizontal lines indicate in each panel the canonical high-mass IMF index, $\alpha_3=2.3$. In this Figure, the remnants of stars with $m>25 \, \rmn{M}_{\odot}$ are assumed to have masses of 10\% of the initial mass of their progenitors whereever this is relevant.}
\label{alphaM}
\end{figure*}

It was shown in Section~\ref{alpha3Mass} that there is no hard evidence for a correlation of the $\Upsilon_{V\rmn{,n}}$ of the UCDs with their mass if UCDs and GCs are separate populations. For this case, it is therefore a useful and good assumption that all UCDs have the same $\Upsilon_{V\rmn{,n}}$ and that deviations from it are due to statistical scatter. Thus, we estimate the $\Upsilon_{V\rmn{,n}}$ of the UCDs and the uncertainty of this value by performing a least-squares fit of eq.~(\ref{eqMUps}) with $A=0$ to the $\Upsilon_{V\rmn{,n}}$ of the UCDs. The best-fitting parameter $B$ then equals to the numerical value of the mean of the $\Upsilon_{V\rmn{,n}}$ of the individual UCDs, $\overline{\Upsilon_{\rmn{UCD}}}$, and is $(4.75 \pm 0.34) \, \rmn{M}_{\odot} \, \rmn{L}_{\odot ,V}^{-1}$ if the age of the UCDs is assumed to be $7 \, \rmn{Gyr}$ and $(4.97 \pm 0.37) \, \rmn{M}_{\odot} \, \rmn{L}_{\odot ,V}^{-1}$ if the age of the UCDs is assumed to be $13 \, \rmn{Gyr}$. $\overline{\Upsilon_{\rmn{UCD}}}$ is shown as the horizontal dotted line in Fig.~\ref{MLratios} and as the vertical dotted line in Fig.~\ref{alphaUps}. The uncertainties of $\overline{\Upsilon_{\rmn{UCD}}}$ are indicated as shaded areas in these figures.

A given value for $\overline{\Upsilon_{\rmn{UCD}}}$ is taken to depend only on $\alpha_3$. This implies that all UCDs have formed with the same (top-heavy) IMF, which can be considered as characteristic for very dense star-forming regions. Note however that even if this assumption is consistent with the available data, it is a simplification because the $\Upsilon_{V\rmn{,n}}$ of the UCDs are \emph{expected} to scatter due to age differences. Furthermore, the suggestion that the IMF is top-heavy in UCDs in comparison to less massive stellar systems is based on the notion that the process of star formation (and thus the IMF) depends on the physical conditions under which it takes place. This implies that there can be as many IMFs as physical conditions under which star formation takes place.

$\overline{\Upsilon_{\rmn{UCD}}}$ can be translated into different expected values for $\alpha_3$, depending on the assumed remnant population in the UCDs. The upper limit to the expected $\alpha_3$ can be obtained from the lower bound of the uncertainty of $\overline{\Upsilon_{\rmn{UCD}}}$, and the lower limit to the expected $\alpha_3$ can be obtained from the upper bound of the uncertainty of $\overline{\Upsilon_{\rmn{UCD}}}$. The values for $\alpha_3$ are listed in Tab.~\ref{alpha3}.

\citet{Mie2006} found that the UCDs in Fornax and the ones in Virgo are (despite their similarity) distinct in their properties, which could indicate a different origin or age for the two groups. Among these distinctive properties is the average $\Upsilon_{V\rmn{,n}}$ of the UCDs, which is clearly higher for the ones in the Virgo Cluster. It is therefore worthwhile to relax the assumption of a common $\Upsilon_{V\rmn{,n}}$ for all UCDs and assume a common $\Upsilon_{V\rmn{,n}}$ only for the UCDs that are bound to the same larger structure (i.e. the Fornax Cluster, the Virgo Cluster or Centaurus~A), although the smaller size of the subsamples decreases their statistical significance. Estimates for $\alpha_3$ for the UCDs in the different subsamples can then be obtained in the same way as for the whole sample. The results are given in Tab.~\ref{alpha3}.

Note that the values obtained for $\alpha_3$ for the UCDs in Fornax have large uncertainties if they are assumed to be $13 \, \rmn{Gyr}$ old. This is because in that case $\overline{\Upsilon_{\rmn{UCD}}}$ for them is quite close to the $\Upsilon_{V\rmn{,n}}$ where $\alpha_3$ asympotically approaches infinity. Consequently $\alpha_3$ is obtained from an interval in $\Upsilon_{V\rmn{,n}}$ where small variations of $\Upsilon_{V\rmn{,n}}$ imply large changes in $\alpha_3$. This applies in particular to the upper limit to $\alpha_3$. Thus, the numbers in question are only of use for giving lower bounds for the high-mass IMF slope, which are obtained at a $\Upsilon_{V\rmn{,n}}$ where the dependency of $\alpha_3$ on $\Upsilon_{V\rmn{,n}}$ is more moderate.

Also note the strong impact of S999 with its extreme $\Upsilon_{V\rmn{,n}}$ (upper-most data-point in Figs.~\ref{MLratios} and~\ref{MUps})on the $\alpha_3$ derived for the UCDs in the Virgo Cluster. The relevance of this particular cluster is evidently much smaller if the whole sample of UCDs is considered. The main results presented in this paper are therefore either not or only mildly affected by this stellar system. In particular, S999 plays no role for deciding whether there is a significant correlation between the $\Upsilon_{V\rmn{,n}}$ of the UCDs and their mass. Such an outlier may be due to recently induced tidal effects \citep{Fel2006}.

\subsubsection{GCs and UCDs as a single population}
\label{single}

Contrary to the case that UCDs and GCs constitute different populations, the slope $A$ is highly significant if UCDs and GCs are a single population. A relation between $\alpha_3$ and mass, $\alpha_3(M)$, can be established by solving eq.~(\ref{eqML2}) for the different $\Upsilon_{V\rmn{,n}}(M)$ corresponding to different masses. The results for this are plotted in Fig.~\ref{alphaM}. The fact that in the mass range of GCs, $\Upsilon_{V\rmn{,n}}(M)$ almost coincides with the model predictions for a SSP with the canonical IMF for an assumed age of $7 \, \rmn{Gyr}$ is purely coincidental. Independent estimates on the ages of MWGCs (which are the bulk of the GCs in the sample used here) are closer to $13 \, \rmn{Gyr}$ than to $7 \, \rmn{Gyr}$ for most of them \citep{Van2000,Sal2002}. However, if an age of $13 \, \rmn{Gyr}$ is assumed for the GCs and UCDs, $\alpha_3(M)$ is not defined for $M\apprle 3 \times 10^6 \, \rmn{M}_{\odot}$ because in this mass range $\Upsilon_{V\rmn{,n}}(M)$ is below the minimum value for which eq.~(\ref{eqML2}) is solvable.

This finding implies that the present-day stellar mass function of the corresponding stellar systems has to be poorer in very low-mass stars than the canonical IMF, since their  $\Upsilon_{V\rmn{,n}}$ cannot be realised in any case if the mass function of their main sequence stars equals the canonical IMF. There are different processes which tend to drive very low-mass stars out of a star-cluster. One of them is dynamical evolution (cf. \citealt{Kru2008a} and \citealt{Kru2008b}). It acts faster the less massive a star cluster is. However, the expected effect on the $M/L_V$ ratio is only small according to \citet{Bau2003}, their figure~14 and \citet{Bor2007}. More relevant would be gas expulsion if the GCs were initially mass-segregated \citep{Mar2008}. In other words, if GCs have formed with the canonical IMF, their stellar mass functions must have changed with time (see also DHK and \citealt{Mie2008b}). The dependency of the stellar mass function on the cluster concentration \citep{DeM2007} indeed suggests this to be the case. Consequently, the assumption of a stellar population only altered by stellar evolution would only be valid for stellar systems with $M \apprge 3 \times 10^6 \, \rmn{M}_{\odot}$.

For a parametrisation of $\alpha_3$ as a function of the mass of a stellar system we suggest
\begin{equation}
\overline{\alpha_3}(M)= \left[ \log_{10} \left( \frac{0.85 (10^{-6} \, M/\rmn{M}_{\odot})^2}{(10^{-6} \, M/\rmn{M}_{\odot})-1} \right) \right]^{-1}+0.42,
\label{eqalpha3}
\end{equation}
for $M \ge 2 \times 10^6 \, \rmn{M}_{\odot}$, and $\overline{\alpha_3}(M)=2.3$ for $M < 2 \times 10^6 \, \rmn{M}_{\odot}$, where $M$ is measured in $\rmn{M}_{\odot}$. $\overline{\alpha_3}(M)$ thus returns the canonical IMF for GCs, which is motivated with the invariance of the IMF in resolved stellar populations \citep{Kro2001,Kro2008,Mar2008}. In the range of massive UCDs, which are the least vulnerable to dynamical evolution, $\overline{\alpha_3}(M)$ is chosen to be roughly the mean of $\alpha_3(M)$ for assumed ages of $7 \, \rmn{Gyr}$ and $13 \, \rmn{Gyr}$ at a NS and BH retention rate of 20\%. Note that the change of $\alpha_3(M)$ in this mass range is only moderate for the two extreme assumptions on the age of the UCDs. In the intermediate mass range from $2 \times 10^6 \, \rmn{M}_{\odot}$ to $\approx 10^7 \, \rmn{M}_{\odot}$, $\overline{\alpha_3}(M)$ is an (in principle arbitrary) interpolation from the low-mass regime to the very high-mass regime. $\overline{\alpha_3}(M)$ is plotted in Fig.~\ref{para}.

\section{discussion}
\label{discussion}

\subsection{Stability of the UCDs}
\label{stability}

\begin{figure}
\includegraphics[scale=0.78]{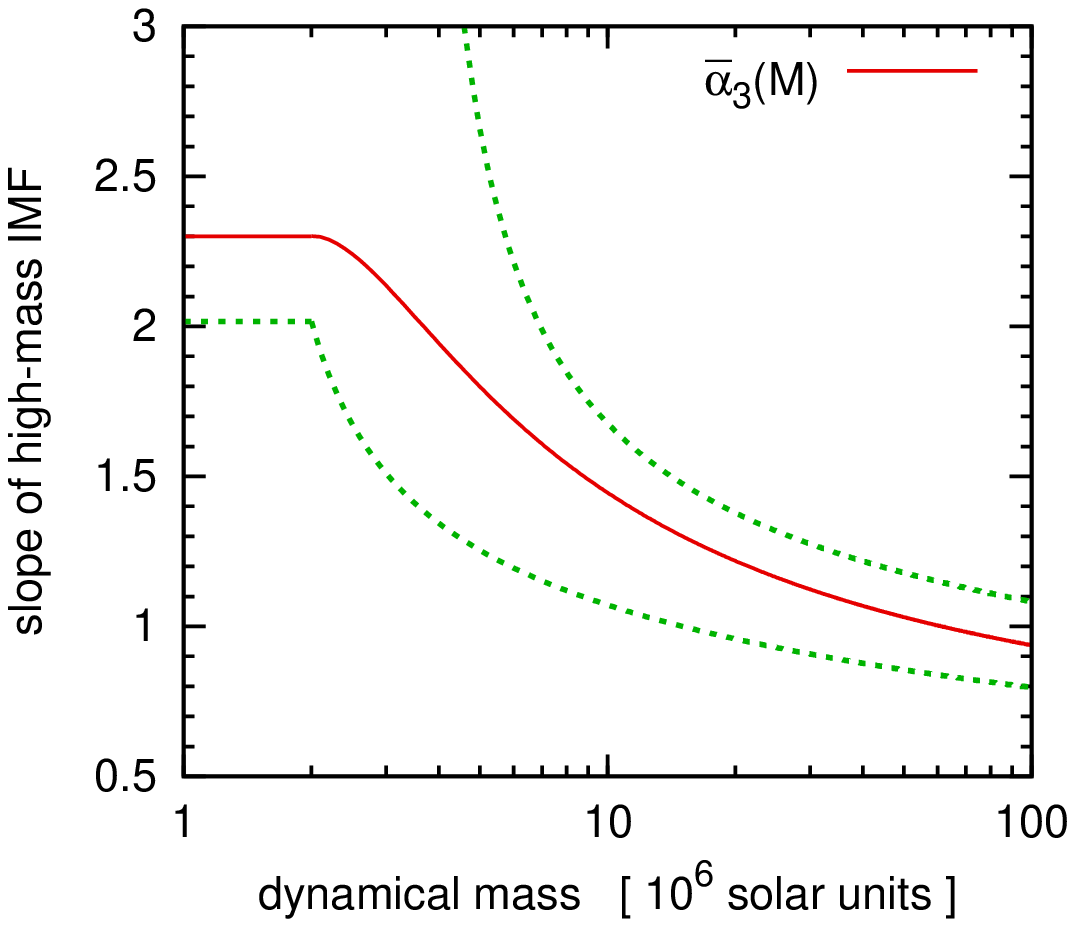}
\caption{The parametrisation of the high mass IMF slope, $\alpha_3$, as a function of the mass of a stellar system, $M$, as given by eq.~(\ref{eqalpha3}). It is indicated by the solid line. Also shown are $\alpha_3(M)$ as found from eq.~(\ref{eqML2}) using eq.~(\ref{eqMUps}) assuming a NS and BH retention rate of 20\% and BHs having 10\% of the initial mass of their progenitor stars. The upper dashed line corresponds to an estimated age of $13 \, \rmn{Gyr}$ and the lower dashed line to an estimated age of $7 \, \rmn{Gyr}$ for the UCDs.}
\label{para}
\end{figure}

\citet{Bau2008} show that for a star cluster with the canonical IMF the combined energy input from all SNe exceeds the binding energy of star clusters with initial masses up to $\approx 10^7 \, \rmn{M}_{\odot}$ (cf. their fig.~3)\footnote{Star clusters less massive than $10^7 \, \rmn{M}_{\odot}$ can survive this energy input because the energy from the SNe is not distributed uniformly on all matter in the cluster.}. This implies that star clusters loose not only most of the matter bound in massive stars, but also their primordial gas in less than $40 \, \rmn{Myr}$ ($\approx 40 \, \rmn{Myr}$ is the time it takes until all massive stars in a SSP have evolved, cf. the grids by \citealt{Sch1992}). Although many UCDs certainly had initial masses higher than $10^7 \rmn{M}_{\odot}$, they also had many more massive stars that evolved into SNe if they formed with IMFs as top-heavy as suggested in this paper. In this case, they would loose an even larger fraction of their initial mass during their early evolution than less massive stellar systems, because of the large mass-faction bound in massive stars. For instance, 23.0\% of the total initial stellar mass of a star cluster is in stars more massive than $8 \, \rmn{M}_{\odot}$ if $\alpha_3=2.3$. This value rises to 73.0\% for $\alpha_3=1.57$ (which is the high-mass IMF slope suggested in Tab.~\ref{alpha3} for $13 \, \rmn{Gyr}$ old UCDs that retain 20\% of their NSs and BHs with $m_{\rmn{BH}}=0.1 m$) and to 93.3\% for $\alpha_3=1.04$ (which is the corresponding value for $7 \, \rmn{Gyr}$ old UCDs).

Observations of star-forming regions in the Milky Way show that only a fraction of available gas is actually converted into stars (e.g. \citealt{Lad2003}). Assuming that this left-over gas is swept out of young star clusters by the radiation and evolution of massive stars, the total mass-loss of the stellar system until the end of massive-star evolution can be written as
\begin{equation}
 M_{\rmn{init}}-M_{\rmn{final}}=M_{\rmn{init}}[1-\rmn{SFE}(1-x)],
\label{Massloss}
\end{equation}
where $M_{\rmn{init}}$ and $M_{\rmn{final}}$ are the initial mass of the stellar system and the final mass of the stellar system respectively (stars and gas), SFE is star formation efficiency of the stellar system and $x$ the mass-fraction of stars with $m>8 \, \rmn{M}_{\odot}$. It is thereby assumed in eq.~(\ref{Massloss}) that the total mass of the remnants remaining in the stellar system is negligible compared to the total initial mass of all stars with $m>8 \, \rmn{M}_{\odot}$. This approximation is well fulfilled for all cases where the mass of the remnants of massive stars is rather small or only a few of them remain inside the stellar system. (These cases correspond to our models i, ii, iii and iv for the composition of the UCDs; see Section~\ref{processed}. That is why the IMFs estimated for the UCDs turn out to be so flat if one of these models is assumed for their composition, even if the difference between the predicted  $M/L$ ratio and the observed $M/L$ ratio is not very large.) Assuming that the star formation efficiency in a UCD is 0.4, the mass-loss within the first $40 \, \rmn{Myr}$ is, according to eq.~\ref{Massloss}, 69.2\% of the total initial mass (stellar and gas) if the high-mass IMF slope of the UCD was $\alpha_3=2.3$ (canonical IMF), but 89.2\% for $\alpha_3=1.57$ and 97.3\% for $\alpha_3=1.04$. (Note that a SFE of 0.4 would be a high value for an open cluster, but not necessarily for a UCD, cf. \citealt{Mur2008}. Note also that we ignore the probably significant loss of stars from the UCD due to the unbinding effect from gas expulsion. We return to this in a follow-up paper.)

The behaviour of a stellar system that loses a large fraction of its initial mass very much depends on the rate of the mass-loss. This behaviour can be characterised by two limiting cases:

\begin{itemize}
\item Rapid mass-loss (i.e. the mass-loss takes place on a timescale shorter or comparable to the crossing time): Arguing with the virial theorem, \citet{Hil1980} finds from analytic estimates that a star cluster dissolves if it loses more than 50\% of its initial mass instantaneously. Using $N$-body integrations, \citet{Boi2003} and \citet{Fel2005} show that the survival of a star cluster is also dependent on the density profile of the star cluster and the velocity distribution of its stars, but a sudden loss of more than 67\% of its initial mass is critical in any case.
\item Adiabatic mass-loss (i.e. the mass-loss is slow enough for the stellar system to stay near virial equilibrium at all times): Adiabatic mass-loss does not unbind the remainder of the star cluster, but inflates it. The change in radius is
\begin{equation}
\frac{r_{\rmn{final}}}{r_{\rmn{init}}}=\frac{M_{\rmn{init}}}{M_{\rmn{final}}},
\label{adiabatic}
\end{equation}
where $r_{\rmn{init}}$ and $M_{\rmn{init}}$ are the radius and mass, respectively, of the stellar system at the beginning of mass-loss and $r_{\rmn{final}}$ and $M_{\rmn{final}}$ the according parameters at the end of mass-loss \citep{Kro2008}.
\end{itemize}

Important numbers for deciding whether the mass-loss from UCDs is rapid or adiabatic are their crossing times, $t_{\rmn{cr}}$, which is defined as $t_{\rmn{cr}}=2r_h/\sigma$, with $r_h$ being the 3D half-mass radius and $\sigma$ being the 3D velocity dispersion of the stellar system \citep{Kro2008}. Furthermore the ratios between half-mass radius and tidal radius, $r_{\rmn{h}}/r_{\rmn{t}}$, decide upon survival of the stellar system. The data published in \citet{Evs2007} and \citet{Hil2007} for the properties of UCDs as they are observed today imply $r_{\rmn{h}}/r_{\rmn{t}}$ well below 0.1 for most of them (sometimes as low as 0.01) and $t_{\rmn{cr}}$ of the order of $1 \, \rmn{Myr}$. This suggests a timespan of the order of $40 \, t_{\rmn{cr}}$ for the timescale for SN-driven mass-loss.

In the grid of $N$-body simulations performed by \citet{Bau2007}, star-clusters are predicted to dissolve if they loose 95\% of their initial mass, even for the most moderate tidal fields ($r_{\rmn{h}}/r_{\rmn{t}}=0.01$) and longest duration for mass-loss (10 $t_{\rmn{cr}}$) they consider. However, if the stellar system looses only 90\% of its initial mass on that timescale, the stellar system may not completely dissolve as long as the tidal field is weak. The mass fraction of the remnant of the star cluster that continues to be gravitationally bound after it has returned to virial equilibrium is then 0.65 for $r_{\rmn{h}}/r_{\rmn{t}}=0.01$ and 0.35 for $r_{\rmn{h}}/r_{\rmn{t}}=0.033$, while the half-mass radius increases to approximately ten times its initial value in both cases. This implies that the stellar density in those systems decreases to less than $10^{-3}$ times its initial value. If also the gas leaving intermediate mass stars as they evolve into WDs is driven out of the UCDs (e.g. through type I SNe), the mass of the UCDs is decreased further to 0.57 of its original value for $\alpha_3=1.57$ and to 0.40 of its original value for $\alpha_3=1.04$. Since this mass-loss would be adiabatic, the according change in radius can be calculated using eq.~(\ref{adiabatic}) and is a factor of 1.75 for $\alpha_3=1.57$ and a factor of 2.48 for $\alpha_3=1.04$. The density would thus be further decreased by a factor $\apprle 10$. At present, the UCDs typically have mean central densities from $10^{2}$ to $10^3 \, \rmn{M}_{\odot} \, \rmn{pc}^{-3}$ (fig.~4 in DHK). A mass-loss of 90\% of the initial mass (stellar and gas) over the first $\approx 40 \, t_\rmn{cr}$ and subsequent adiabatic mass-loss through the evolution of intermediate-mass stars would therefore suggest initial central densities of at least $10^{6}$ to $10^{7} \, \rmn{M}_{\odot} \, \rmn{pc}^{-3}$. This corresponds to $3.9\times 10^5$ to $3.9\times 10^6$ stars per $\rmn{pc}^{3}$ for $\alpha_3=1.57$ and $8.4\times 10^4$ to $8.4\times 10^5$ stars per $\rmn{pc}^{3}$ for $\alpha_3=1.04$, the fraction of stars more massive than $8 \, \rmn{M}_{\odot}$ among them being 5.5\% for $\alpha_3=1.57$ and 23.7\% for $\alpha_3=1.04$. Such systems would thus have had extensions similar to GCs (i.e. $r_\rmn{h}$ of a few pc). Typical total initial stellar masses would be some $10^{7} \, \rmn{M}_{\odot}$, implying a population of $\approx 10^6$ stars with $m \ge 8 \, \rmn{M}_{\odot}$ for $\alpha_3=1.57$ as well as for $\alpha_3=1.04$.

These numbers underline the extreme nature of UCDs. Their stability seems questionable if they would have formed with a top-heavy IMF and their contemporary structural parameters. However, the smaller extensions and higher masses the UCDs must have had before evolutionary processes set in imply that the conditions for adiabatic mass-loss were fulfilled much better at that time. However, also if mass-loss from a UCD with a top-heavy IMF was adiabatic at all times, and its stability was therefore not threatened, eq.~\ref{Massloss} still implies an enormous inflation nd decrease of density for it.

The observation of UCDs today therefore does not contradict a formation scenario with a very top-heavy IMF for them. A more detailed, numerical study of this issue will be provided in a follow-up paper.

\subsection{The Star formation rate in UCDs at their formation}
\label{starformation}

The notion that UCDs might be the most massive star clusters implies that they formed from a collapsing molecular cloud. Star formation within the cloud is thought to set in as soon as a certain density is reached, which is according to \citet{Kaw1998} at a column density in excess of $1.6\times10^{21} \, N(H_2)\, \rmn{cm}^{-2}$. Defining the size, $R$, of a cloud as $(S/\pi)^{0.5}$ where $S$ is the total cloud surface area, gives typical sizes of $3 \, \rmn{pc}$ for the clouds in the sample of \citet{Kaw1998}. This corresponds to a mean density, $\overline{\rho}$, of $\approx 4 \, \rmn{M}_{\odot} \, \rmn{pc}^{-3}$ as the criterion for the onset of star formation. It then proceeds rapidly and is completed within a timescale of the order of a free fall time \citep{Elm2000,Har2001}. For spherically symmetric matter distributions, the free fall time, $t_{\rmn{ff}}$, is given as $t_{\rmn{ff}}=(3\pi/32G\overline{\rho})^{0.5}$, where $G$ is the gravitational constant. Note the independence of $t_{\rmn{ff}}$ on the total mass. The assumption of spherical symmetry for star-forming gas clouds thus leads to a time scale of $4 \, \rmn{Myr}$ on which star formation takes place, whereby the bulk of the stars may form on an even shorter time scale (for instance, 80\% within $1 \, \rmn{Myr}$ in the Orion Nebula Cluster, \citealt{Pro1994}). If applied to the UCDs, this suggests that their stellar populations formed with star formation rates of $\approx 10 \, \rmn{M}_{\odot} \, \rmn{yr}^{-1}$ to $\approx 100 \, \rmn{M}_{\odot} \, \rmn{yr}^{-1}$, depending on the mass of the UCD. Given that most if not all stars are formed in star clusters \citep{Lad2003} and that the time scale for star cluster formation appears to be independent of the mass of the cluster, these star formation rates would be the highest ever to be found in a single star formation event.

\section{Summary and Conclusions}
\label{conclusions}

It was shown in previous papers that the dynamical $M/L_V$ ratios, $\Upsilon_{V}$, of compact stellar systems more massive than $2 \times 10^6 \, \rmn{M}_{\odot}$ are not consistent with the predictions from simple stellar population models, if the canonical IMF is assumed for star formation in them. Out of the possible explanations for this result (top-heavy IMF, bottom-heavy IMF, dark matter, inaccuracy of the SSP models), the notion of a top-heavy initial stellar mass function (IMF) in dense star-forming regions seems especially attractive.

With this motivation, we quantify by how much the IMF in a sample of massive compact stellar systems (referred to as UCDs) has to deviate from the canonical IMF in the intermediate and high mass part for the modelled $\Upsilon_{V}$ to agree with the observed ones. The model constructed for this accounts for the different metallicities of the UCDs. Several combinations of assumptions concerning age ($7 \, \rmn{Gyr}$ or $13 \, \rmn{Gyr}$) and the amount of processed material with very high $\Upsilon_{V}$ retained by the UCDs besides white dwarfs (no remnants of massive stars; 20\% of the remnants of massive stars; all remnants of massive stars; all material processed by burnt-out stars) are considered. The IMF of the UCDs is taken to be a three-part power-law that equals the canonical IMF below an initial stellar mass of $1 \, \rmn{M}_{\odot}$. The exact upper mass limit of the IMF ($m_{\rmn{max}}=100 \, \rmn{M}_{\odot}$ or $m_{\rmn{max}}=150 \, \rmn{M}_{\odot}$) turns out to have a negligible impact on the results.

Assuming that all UCDs have the same normalised $M/L$ ratio (which is justifiable considering the uncertainties of their $\Upsilon_{V}$) and that the processed material retained by the UCDs are all white dwarfs and 20\% of the remnants of massive stars, our model suggests a high-mass IMF slope, $\alpha_3$, of $\approx 1.6$ if the UCDs are $13 \, \rmn{Gyr}$ old (i.e. almost as old as the Universe) or $\approx 1.0$ if the UCDs are $7 \, \rmn{Gyr}$ old. If the UCDs were assumed to have formed with the canonical IMF, their $\Upsilon_{V}$ would only be explainable if they contain significant amounts of non-baryonic dark matter or dense interstellar gas. Note that there would need to be some mechanism that inhibits on-going star formation in this case.

The $\Upsilon_{V}$ of the UCDs in the Fornax cluster tend to be lower than the ones of the other UCDs. If the Fornax UCDs are assumed to be $13 \, \rmn{Gyr}$ old they have a normal $\Upsilon_{V}$ and consequently not at a top-heavy IMF. Assuming that the discrepancy between the prediction of SSP models with the canonical IMF and the observed $M/L_V$ ratios of the Fornax UCDs as high as for the Virgo UCDs suggests an ages around $7 \, \rmn{Gyr}$ for the Fornax UCDs if the Virgo UCDs are taken to be $13 \, \rmn{Gyr}$ old \citep{Mie2008b}.

The dependency of $\alpha_3$ on the normalised $M/L_V$ ratio, $\Upsilon_{V\rmn{,n}}$, established in eq.~(\ref{eqML2}), can be translated into a dependency of $\alpha_3$ on the mass of the stellar system, $M$. This is done using the increase of $\Upsilon_{V\rmn{,n}}$ with $M$ formulated in eq.~\ref{eqMUps} and shown in Fig.~\ref{MUps}. A possible parametrisation of this dependency is given in eq.~(\ref{eqalpha3}) and plotted in Fig.~\ref{para}.

The mass-loss due to the evolution of massive stars may reach 90\% of the initial stellar mass of a star cluster for very top-heavy IMFs, even if primordial gas expulsion is not considered. The survival of the UCDs seems not to be threatened by the mass loss implied by the evolution of the stars alone, as this mass loss would be adiabatic. However, the radiation and evolution of massive stars also drives the expulsion of the primordial gas. The timescale on which this process takes place is critical for the survival of the UCD. This is an issue deserving further study. In any case, the results in Sections~\ref{stability} and~\ref{starformation} suggest that UCDs formed with likely central stellar densities of $10^{6}$ to $10^{7} \, \rmn{M}_{\odot} \, \rmn{pc}^{-3}$ and possible star formation rates of $\approx 10 \, \rmn{M}_{\odot} \, \rmn{yr}^{-1}$ to $\approx 100 \, \rmn{M}_{\odot} \, \rmn{yr}^{-1}$. These are among the most extreme sites of star formation.

\section*{Acknowledgements}
The authors wish to thank the referee for making comments that improved the paper.
JD acknowledges support through DFG grant KR1635/13.

\bibliographystyle{mn2e}
\bibliography{paper12}

\appendix

\section[]{The total mass of the remnants}
\label{Equations}

We explicitly note the terms with $m>m_{\rmn{to}}$ that arise from the integration of the right hand side of eq.~(\ref{eqMrem2}), if eq.~(\ref{eqMrem1}) is inserted for $m_{\rmn{rem}}(m)$.

The contribution of the white dwarfs to $M_{\rmn{m}}$, $M_{\rmn{m,WD}}$, can be written as
\begin{equation}
\begin{split}
\frac{M_{\rmn{m,WD,v}}}{\rmn{M}_{\odot}}= & \frac{0.109 \, k_3}{2-\alpha_3} \times (8^{2-\alpha_3}-m_{\rmn{to}}^{2-\alpha_3}) \\
+ & \frac{0.394 \, k_3}{1-\alpha_3} \times (8^{1-\alpha_3}-m_{\rmn{to}}^{1-\alpha_3}),
\end{split}
\label{eqWDv}
\end{equation}
with the masses of the stars in Solar units. $m_{\rmn{to}}$ was argued to be $\approx 1 \, \rmn{M}_{\odot}$ in Section \ref{SSP}.

The contribution of neutron stars to $M_{\rmn{m}}$, $M_{\rmn{m,NS}}$, can be written as
\begin{equation}
\frac{M_{\rmn{m,NS}}}{\rmn{M}_{\odot}}= \frac{1.35 \, k_3}{1-\alpha_3}\times ( 25^{1-\alpha_3}-8^{1-\alpha_3}).
\label{eqNS}
\end{equation}

The contribution of the remnants of stars with initial masses higher than $25 \, \rmn{M}_{\odot}$ to $M_{\rmn{m}}$, $M_{\rmn{m,BH}}$, can be written as
\begin{equation}
\frac{M_{\rmn{m,WD,v}}}{\rmn{M}_{\odot}}= \frac{0.1 \, k_3}{2-\alpha_3} \times (m_{\rmn{max}}^{2-\alpha_3}-25^{2-\alpha_3}).
\end{equation}

\label{lastpage}

\end{document}